\documentclass[a4paper,11pt]{article}
\usepackage{color}
\usepackage{jinstpub}
\usepackage{graphicx}
\usepackage{subfigure}
\usepackage{siunitx}
\usepackage{amsmath}
\usepackage{lineno}
\usepackage{booktabs}
\usepackage{soul}
\usepackage{url}
\usepackage[numbers,sort&compress]{natbib}

\title{Light yield and field dependence measurement in PandaX-II dual-phase xenon detector}
\collaboration{PandaX Collaboration}

\author[a]{Zhou Huang}
\author[a]{Abdusalam Abdukerim}
\author[a]{Zihao Bo}
\author[a]{Wei Chen}
\author[a,q]{Xun Chen}
\author[g]{Yunhua Chen}
\author[m]{Chen Cheng}
\author[h,i]{Yunshan Cheng}
\author[k]{Xiangyi Cui}
\author[n]{Yingjie Fan}
\author[a]{Deqing Fang}
\author[o]{Changbo Fu}
\author[f]{Mengting Fu}
\author[b,e]{Lisheng Geng}
\author[a]{Karl Giboni}
\author[a]{Linhui Gu}
\author[g]{Xuyuan Guo}
\author[a]{Ke Han}
\author[a]{Changda He}
\author[g]{Jinrong He}
\author[a]{Di Huang}
\author[p]{Yanlin Huang}
\author[q]{Ruquan Hou}
\author[j]{Xiangdong Ji}
\author[l]{Yonglin Ju}
\author[a]{Chenxiang Li}
\author[g]{Mingchuan Li}
\author[l]{Shu Li}
\author[k]{Shuaijie Li}
\author[c,d]{Qing Lin}
\author[a,k,q,1]{Jianglai Liu\note{Spokesperson.}}
\author[h,i]{Xiaoying Lu}
\author[f]{Lingyin Luo}
\author[a]{Wenbo Ma}
\author[o]{Yugang Ma}
\author[f]{Yajun Mao}
\author[a,q]{Yue Meng}
\author[h,i]{Nasir Shaheed}
\author[a]{Xuyang Ning}
\author[g]{Ningchun Qi}
\author[a]{Zhicheng Qian}
\author[h,i]{Xiangxiang Ren}
\author[g]{Changsong Shang}
\author[b]{Guofang Shen}
\author[a]{Lin Si}
\author[g]{Wenliang Sun}
\author[j]{Andi Tan}
\author[a,q]{Yi Tao}
\author[h,i]{Anqing Wang}
\author[h,i]{Meng Wang}
\author[o]{Qiuhong Wang}
\author[a,r]{Shaobo Wang}
\author[f]{Siguang Wang}
\author[m]{Wei Wang}
\author[l]{Xiuli Wang}
\author[a,q,k]{Zhou Wang}
\author[m]{Mengmeng Wu}
\author[a]{Weihao Wu}
\author[a]{Jingkai Xia}
\author[j]{Mengjiao Xiao}
\author[m]{Xiang Xiao}
\author[k]{Pengwei Xie}
\author[a]{Binbin Yan}
\author[p]{Xiyu Yan}
\author[a]{Jijun Yang}
\author[a]{Yong Yang}
\author[n]{Chunxu Yu}
\author[h,i]{Jumin Yuan}
\author[a]{Ying Yuan}
\author[j]{Dan Zhang}
\author[a]{Minzhen Zhang}
\author[g]{Peng Zhang}
\author[a]{Tao Zhang}
\author[a]{Li Zhao}
\author[p]{Qibin Zheng}
\author[g]{Jifang Zhou}
\author[a,2]{Ning Zhou\note{Corresponding author.}}
\author[b]{Xiaopeng Zhou}
\author[g]{Yong Zhou}

\affiliation[a]{School of Physics and Astronomy, Shanghai Jiao Tong University, MOE Key Laboratory for Particle Astrophysics and Cosmology, Shanghai Key Laboratory for Particle Physics and Cosmology, Shanghai 200240, China}
\affiliation[b]{School of Physics, Beihang University, Beijing 100191, China}
\affiliation[c]{State Key Laboratory of Particle Detection and Electronics, University of Science and Technology of China, Hefei 230026, China}
\affiliation[d]{Department of Modern Physics, University of Science and Technology of China, Hefei 230026, China}
\affiliation[e]{International Research Center for Nuclei and Particles in the Cosmos \& Beijing Key Laboratory of Advanced Nuclear Materials and Physics, Beihang University, Beijing 100191, China}
\affiliation[f]{School of Physics, Peking University, Beijing 100871, China}
\affiliation[g]{Yalong River Hydropower Development Company, Ltd., 288 Shuanglin Road, Chengdu 610051, China}
\affiliation[h]{Research Center for Particle Science and Technology, Institute of Frontier and Interdisciplinary Science, Shandong University, Qingdao 266237, Shandong, China}
\affiliation[i]{Key Laboratory of Particle Physics and Particle Irradiation of Ministry of Education, Shandong University, Qingdao 266237, Shandong, China}
\affiliation[j]{Department of Physics, University of Maryland, College Park, Maryland 20742, USA}
\affiliation[k]{Tsung-Dao Lee Institute, Shanghai Jiao Tong University, Shanghai, 200240, China}
\affiliation[l]{School of Mechanical Engineering, Shanghai Jiao Tong University, Shanghai 200240, China}
\affiliation[m]{School of Physics, Sun Yat-Sen University, Guangzhou 510275, China}
\affiliation[n]{School of Physics, Nankai University, Tianjin 300071, China}
\affiliation[o]{Key Laboratory of Nuclear Physics and Ion-beam Application (MOE), Institute of Modern Physics, Fudan University, Shanghai 200433, China}
\affiliation[p]{School of Medical Instrument and Food Engineering, University of Shanghai for Science and Technology, Shanghai 200093, China}
\affiliation[q]{Shanghai Jiao Tong University Sichuan Research Institute, Chengdu 610213, China}
\affiliation[r]{SJTU Paris Elite Institute of Technology, Shanghai Jiao Tong University, Shanghai, 200240, China}

\emailAdd{nzhou@sjtu.edu.cn}

\abstract{
The dual-phase xenon time projection chamber (TPC) is one of the most sensitive detector technology 
for dark matter direct search, 
where the energy deposition of incoming particle can be converted into photons and electrons 
through xenon excitation and ionization. 
The detector response to signal energy deposition varies 
significantly with the electric field in liquid xenon. 
We study the detector's light yield and its dependence 
on the electric field in the PandaX-II dual-phase detector 
containing 580~kg liquid xenon in the sensitive volume. 
From our measurements, the light yield at electric fields
from 0~V/cm to 317~V/cm is obtained for energy depositions up to 236~keV.
}

\arxivnumber{2111.01492}
\begin{document}

\maketitle

\section{Introduction}

In recent years, dual-phase xenon detectors have been driving the dark matter direct detection sensitivity by several orders of magnitudes~\cite{xiao2015panda,Tan_2016,Cui:2017nnn,wang2020results,luxFinal,xenon1tFinal,meng2021p4}. 
When an incident dark matter particle scatters with liquid xenon target, certain amount of energy will be deposited in the detector and prompt scintillation light ($S1$) and ionized electrons are produced.
The scintillation light with a wavelength of 175~nm~\cite{xenon175wavelength} travels in the liquid or gaseous xenon, 
gets reflected by polytetrafluoroethylene (PTFE) reflectors 
and is eventually collected by photo-multiplier tubes (PMTs) at the top and bottom of the detector. 
The ionized electrons move upward to the liquid xenon surface under the electric field.
They are subsequently extracted into the gaseous xenon and converted into secondary scintillation light ($S2$) by a stronger extraction electric field. 
In the energy region of a few tens keV to several hundreds keV, the signals are triggered by $S1$. 
The following $S2$ is paired with the triggered $S1$ and they form a complete physics event. 
Based on the $S2$ light pattern on the top PMT array and the time difference between $S1$ and $S2$ signals (drift time), 
3-D position information of a physics event can be reconstructed~\cite{Tan_2016}. 
The corresponding energy deposition is reconstructed using $S1$ and $S2$ with the formula
\begin{equation}\label{eqn:Erec}
   E_{\text{dep} } = {\rm 0.0137~keV} \left( \frac{S1}{\text{PDE}} + \frac{S2}{ \text{EEE} \times \text{SEG}} \right),
\end{equation}
where PDE, EEE, and SEG are photon detection efficiency, electron extraction efficiency, and single electron gain, respectively. 

The integrated charge of $S1$ produced by energy deposition in liquid xenon is one of the key parameters of the xenon detector response. We can define the xenon light yield as the number of photons per unit energy deposition (keV),
\begin{equation}\label{eqn:LyFormula}
LY = \frac{S1}{\text{PDE} \times E_{\text{dep} }}~,
\end{equation}
where PDE $(12.0 \pm 0.5\%)$ follows Ref.~\cite{wang2020results}.

For a given energy deposition, there are several factors affecting the light yield. 
Part of the energy of nuclear recoil (NR) events will transfer to the atom motion~\cite{Plante:2011hw}.
This effect yields less light signal for NR,
and that did not show in electron recoil (ER).
The electron-ion recombination probability also depends on the strength of drift electric field.
A stronger field is expected to give a smaller light yield. 
Recent large-scale dual-phase xenon experiments include PandaX-II with 580~kg xenon~\cite{Cui:2017nnn,wang2020results},
XENON1T~\cite{xenon1tFinal} with 2~tonne and LUX with 180~kg~\cite{luxFinal}. 
For the optimization of detector operation, 
the three experiments chose different high voltages on the detector electrodes.
The corresponding strengths of drift electric fields are 317~V/cm (PandaX-II), 
81~V/cm (XENON1T) and 180~V/cm (LUX).
For the low energy WIMP search region, Ref.~\cite{ma2020internal,yan2021determination} (PandaX-II),
Ref.~\cite{aprile2018simultaneous,2019xenon1tmodel} (XENON1T) 
and Ref.~\cite{osti_1257742,akerib2017signal,lux2016dd} (LUX)
have described the detector responses 
and signal yields of these three experiments.

In this paper, we report a measurement of xenon light yield 
under various electric fields using PandaX-II detector, 
from 0~V/cm to 317~V/cm. 
The energy points include 9.4~keV, 32.1~keV, 41.5~keV, 164~keV and 236~keV. 
The results are compared with other experiments 
and also a widely used model, NEST model~\cite{Szydagis_2021}.

\section{Monoenergetic event selection}
In order to measure the light yields in different energy regions, 
activated xenon states and $^{83\text{m}}$Kr calibration sources 
were injected into the PandaX-II detector in February 2019, 
after scientific data taking accomplished~\cite{wang2020results},
yielding monoenergetic ER data covering energy from 9.4~keV to 236~keV.
These ER events were taken under four electric field conditions, 317~V/cm, 180~V/cm, 81~V/cm and 0~V/cm,
as summarized in Table~\ref{tab:cali-duration}.
For non-zero field conditions, 
identical quality cuts are inherited from Ref.~\cite{wang2020results} and applied to the data.
While for the zero field condition, only $S1$ relevant quality cuts are applied.
To select these monoenergetic data as clean as possible, the fiducial volume (FV) cut is also applied.
For 317~V/cm condition, the FV is the same with Run 11 FV in Ref.~\cite{wang2020results},
i.e., $50~\mu {\rm s} < $ drift time $< 350~\mu {\rm s}$ and $R^2 < 7.2 \times 10^4~\rm{mm}^2$.
For other non-zero field conditions,
they share the same $R^2$ range cut and the drift time confinements are scaled by the maximum drift times.
Specially, $R^2$ range is unlimited in zero field data due to the lack of $S2$ signal
and an equivalent drift time cut is developed and applied,
which is discussed in section~\ref{sec:kr83m}.

\begin{table}[htbp]
\centering
\caption{\label{tab:cali-duration}Calibration data taking conditions.}
\smallskip
\begin{tabular}{|c|cccc|}
\hline
Drift field strength            &   317~V/cm                &   180~V/cm                &   81~V/cm     & 0~V/cm        \\
\hline
Duration($^{83\text{m}}$Kr)     &   17 hours                &   204 hours               &      272 hours           & 174 hours     \\
Event number($^{83\text{m}}$Kr) &   $\sim 6 \times 10^4$    & $\sim 4.5 \times 10^5$    &    $\sim 6 \times 10^5$  & $\sim 6 \times 10^5$ \\
Duration(Act. Xe)               &   190 hours               &   204 hours               &      272 hours           &    \textbf{---} \\
Event number(Act. Xe)           &   $\sim 5 \times 10^3$    &   $\sim 4 \times 10^3$  &   $\sim 4 \times 10^3$ &    \textbf{---}\\
Maximum drift time              &   360~$\mu s$             &   387~$\mu s$             &       435~$\mu s$  & \textbf{---} \\
\hline
\end{tabular}
\end{table}

\subsection{Activated xenon}
%
One bottle of xenon was stored on the ground surface and exposed to cosmic ray for around three months,
containing approximately 18 kilograms of xenon. 
This bottle of xenon was injected into the PandaX-II detector through the circulation system. 
Due to the neutron activation, natural xenon can be transformed to radioisotopes, 
like $^{127}$Xe, $^{129\text{m}}$Xe and $^{131\text{m}}$Xe.
They produce monoenergetic ER events in the detector, of which 164~keV originates from the $^{131\text{m}}$Xe de-excitation and 236~keV is from both  $^{129\text{m}}$Xe and $^{127}$Xe contributions~\cite{akerib2015radiogenic}.

\begin{figure}[htbp]
    \centering
    \subfigure[]{
        \label{fig:ActXeS2S1}
        \includegraphics[width=0.42\textwidth]{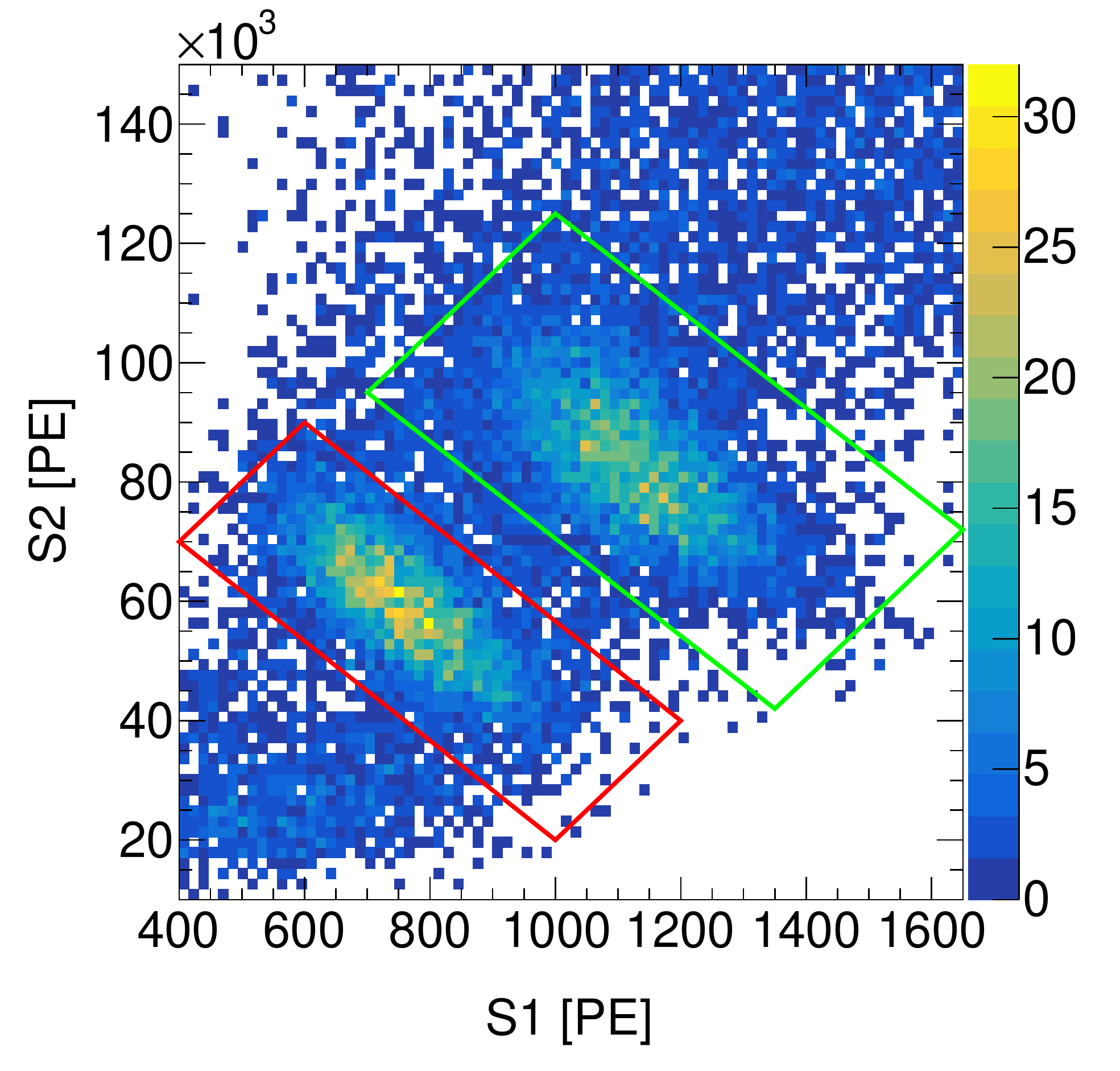}
    }
    \subfigure[]{
        \label{fig:ActXeReduced}
    \includegraphics[width=0.42\textwidth]{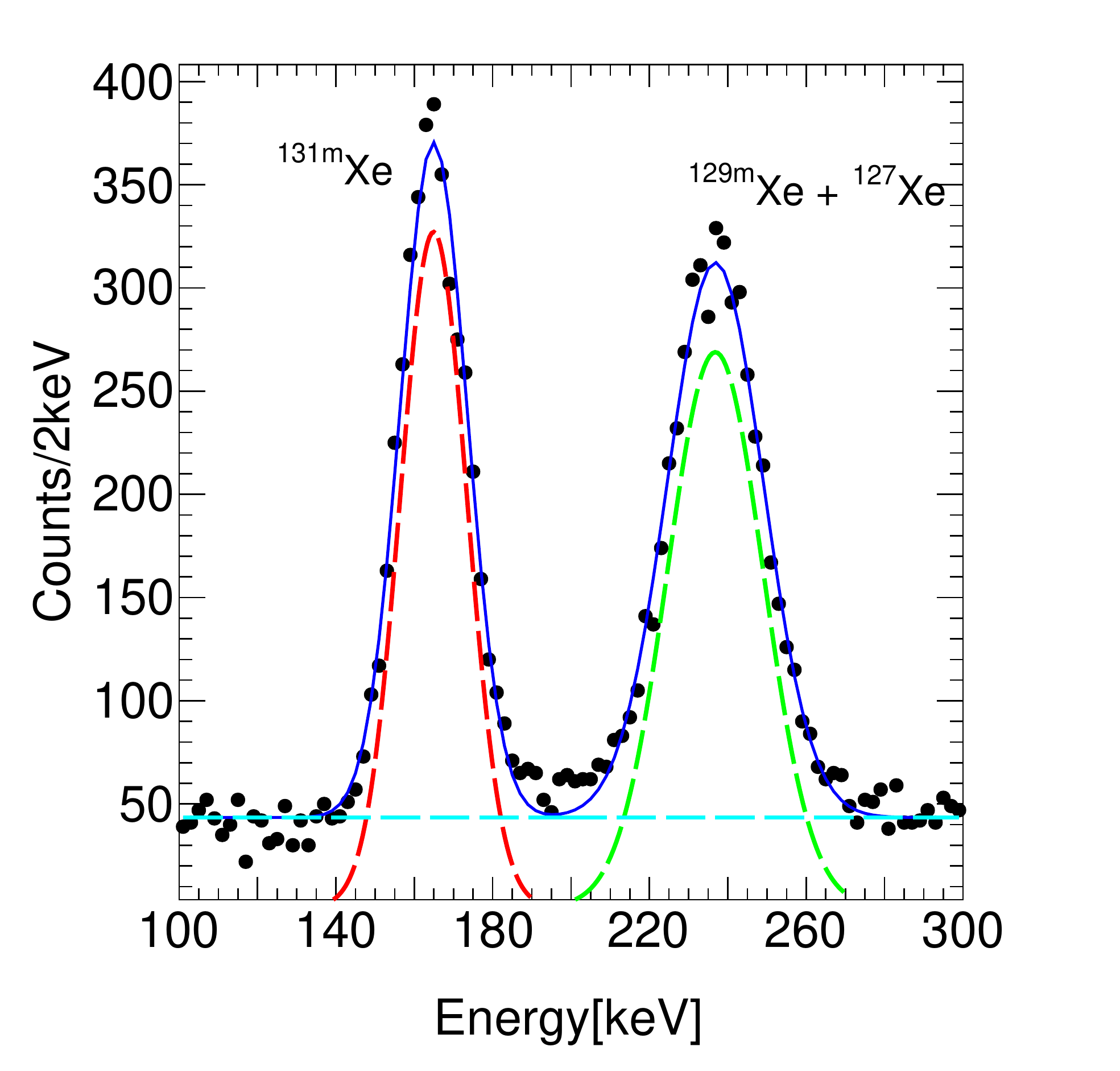}
    }
    \caption{
    (a)Activated xenon event selection in the $S1$-$S2$ distribution. 
    Two anti-correlated contours of the $S1$ and $S2$ signals represent 164~keV events (left red box) and 236~keV events (right green box); (b)Activated xenon energy deposition spectrum. 
    Two major energy peaks are distinguishable and identified at 164~keV (dashed red) and 236~keV (dashed green). 
    Also the background (dashed cyan) is modeled by a linear function.
    }
\end{figure}

From the energy reconstruction formula, monoenergetic events are expected to have an anti-correlation relationship between $S1$ and $S2$, as shown in the $S1$-$S2$ distribution of activated xenon events in Fig.~\ref{fig:ActXeS2S1}, where the two distinguishable ellipse-shaped contours are from 164~keV and 236~keV gammas, respectively. Based on this relationship, we select those monoenergetic events. The reconstructed energy spectrum shows the two energy peaks clearly as well (see Fig.~\ref{fig:ActXeReduced}).

We should note that such a data selection approach is only valid for non-zero drift electric field,
because the ionized electrons would not drift to the gaseous xenon region 
and produce the $S2$ signature if the drift electric field is not applied. 
For the zero electric field case, 
it is difficult to identify the unpaired $S1$ signal of activated xenon 
from the background.
Instead, we rely on the $^{83\text{m}}$Kr isotopes 
whose characteristic decay mode provides a unique signature 
to select the $S1$ signals, as described in section~\ref{sec:kr83m}.

\subsection{\texorpdfstring{$^{83\text{m}}$}{}Kr}
\label{sec:kr83m}
$^{83\text{m}}$Kr is an effective short-lived calibration source, 
which is usually prepared through $^{83}$Rb decay with a half-life 
of 86.2 days~\cite{kr83myale,kr83mzurich,LUXkr83m,zhangdan83mKr}. 
The decay product $^{83\text{m}}$Kr has $J^\pi = \frac{1}{2} ^-$, and de-excites to a $J^\pi = \frac{7}{2} ^+$ state and then the ground state, releasing 32.1~keV and 9.4~keV energy, respectively.
The half-life of the first decay is 1.8 hours, and that of the second is only 156.9~ns~\cite{DataSheet83}. 

\begin{figure}[htbp]
    \centering
    \subfigure[]{
    \includegraphics[width=0.42\textwidth]{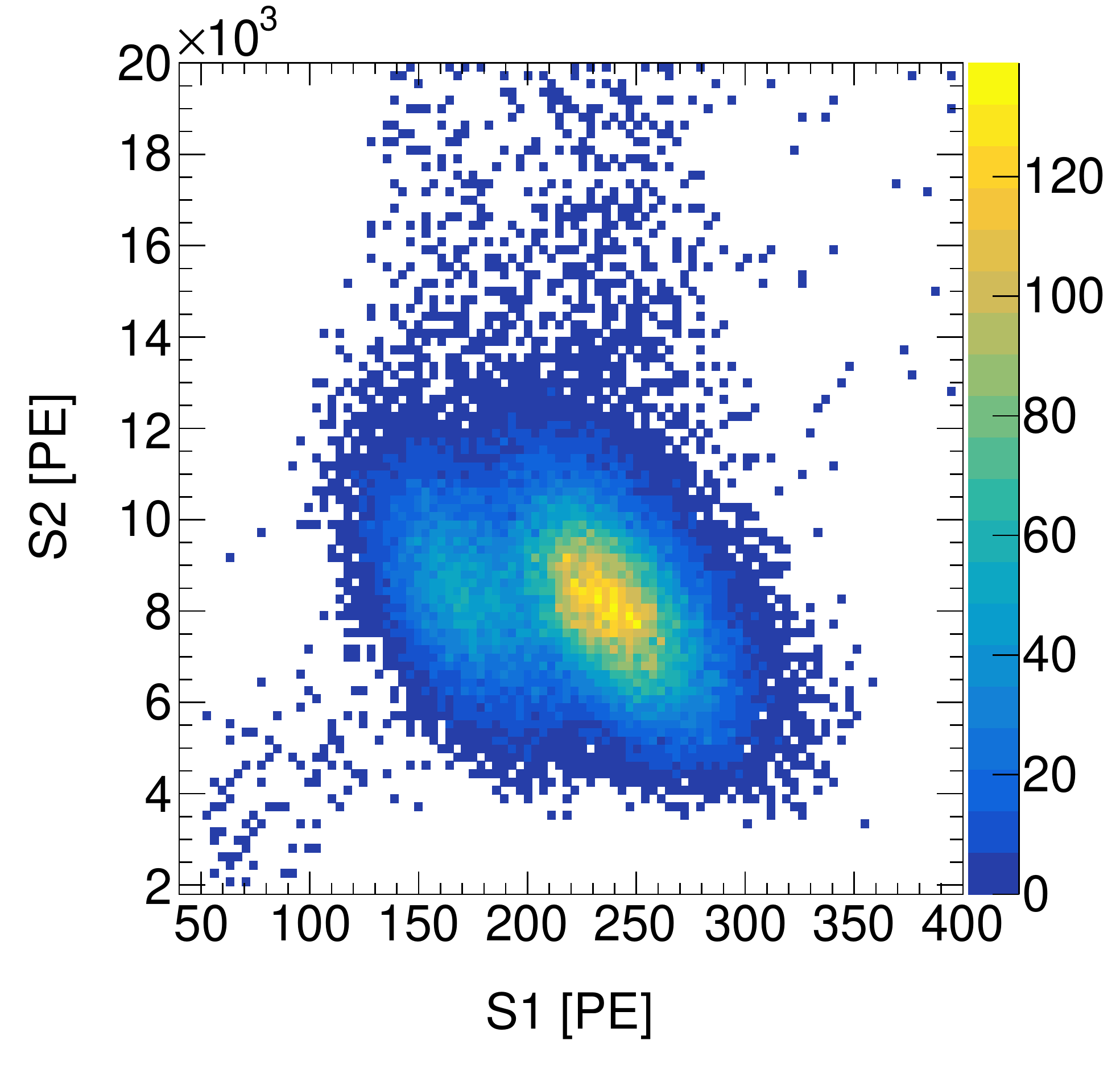}
    \label{fig:kr83mS2S1}
    }
    \subfigure[]{
    \includegraphics[width=0.42\textwidth]{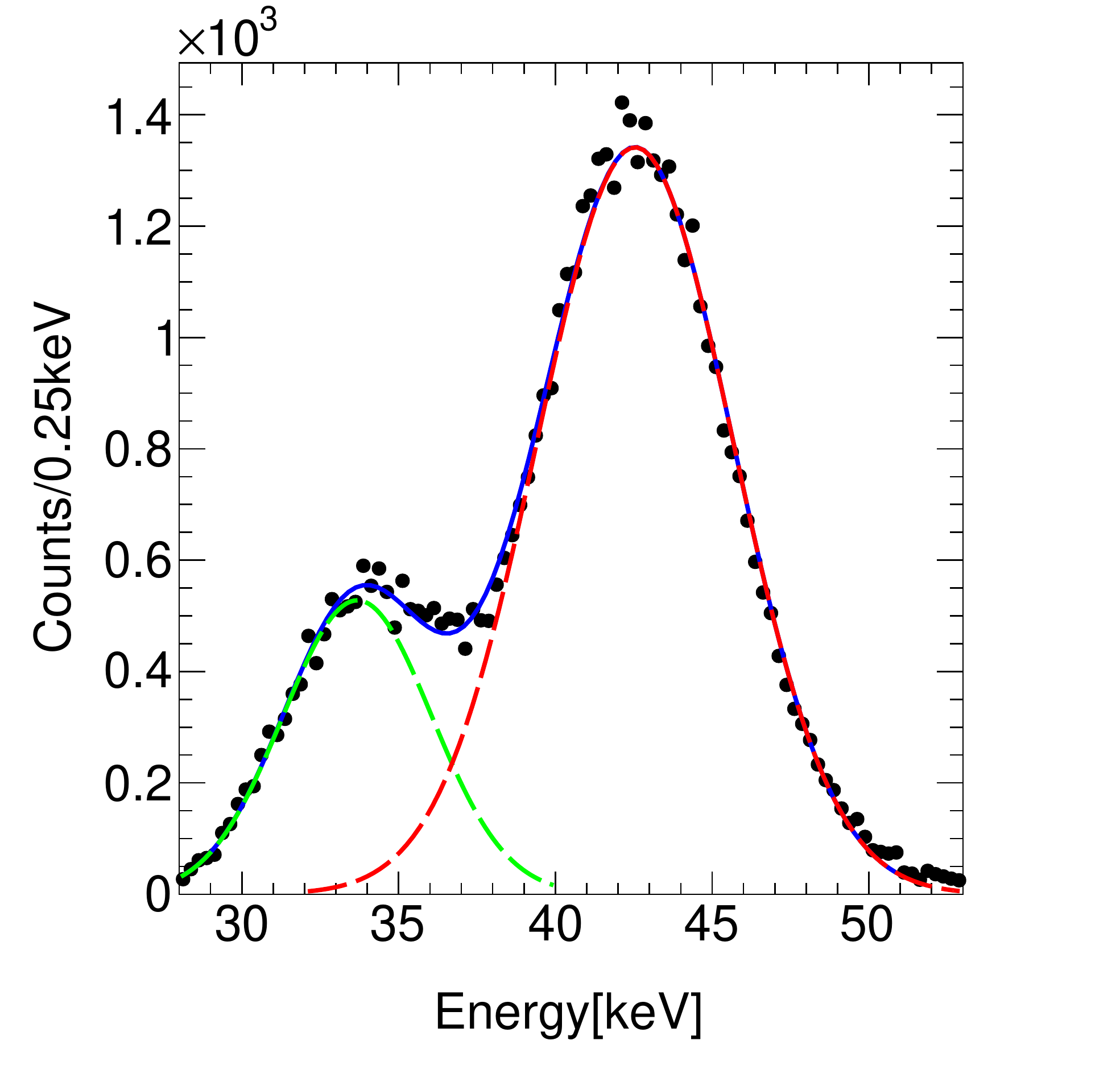}
    \label{fig:kr83mSpec}
    }
    \caption{ (a) $^{83\text{m}}$Kr event $S1$-$S2$ distribution. 
    The $S1$ in this figure represents either two $S1$s combination or 32.1~keV $S1$ only. 
    The two $S2$s are overlapping in the waveform. 
    (b) $^{83\text{m}}$Kr energy deposition spectrum. 
    The major peak (dashed red) is the total cascade decay energy when the two $S1$s from 32.1~keV and 9.4~keV are merged into one. 
    The minor peak (dashed green) corresponds to a lower energy with 9.4~keV $S1$ lost.}
\end{figure}

The typical $S2$ signal waveform width is at a level of microseconds,  so the two sets of ionized electrons from $^{83\text{m}}$Kr cascade decays are overlapping with each other and unable to be separated. For prompt scintillation light with a width of several nanoseconds, there is probability to observe two $S1$ signals in the detector~\cite{wang2020results,kr83myale,kr83mzurich,LUXkr83m,zhangdan83mKr}. Figure~\ref{fig:kr83mS2S1} shows the $^{83\text{m}}$Kr event $S1$-$S2$ distribution, most of the events have two $S1$s merged as one, but some of them have smaller $S1$ corresponding to 32.1~keV decay. 
It indicates that there is another smaller $S1$ from 9.4~keV decay following it. The reconstructed energy spectrum shows a similar behavior (see Fig.~\ref{fig:kr83mSpec}). Therefore, from $^{83\text{m}}$Kr cascade decay, there exist $S1$ signals corresponding to three energy points: 9.4~keV, 32.1~keV and 41.5~keV, which can be selected to study the light yields.

\begin{figure}[htbp]
    \centering
    \subfigure[]{
    \includegraphics[width=0.42\textwidth]{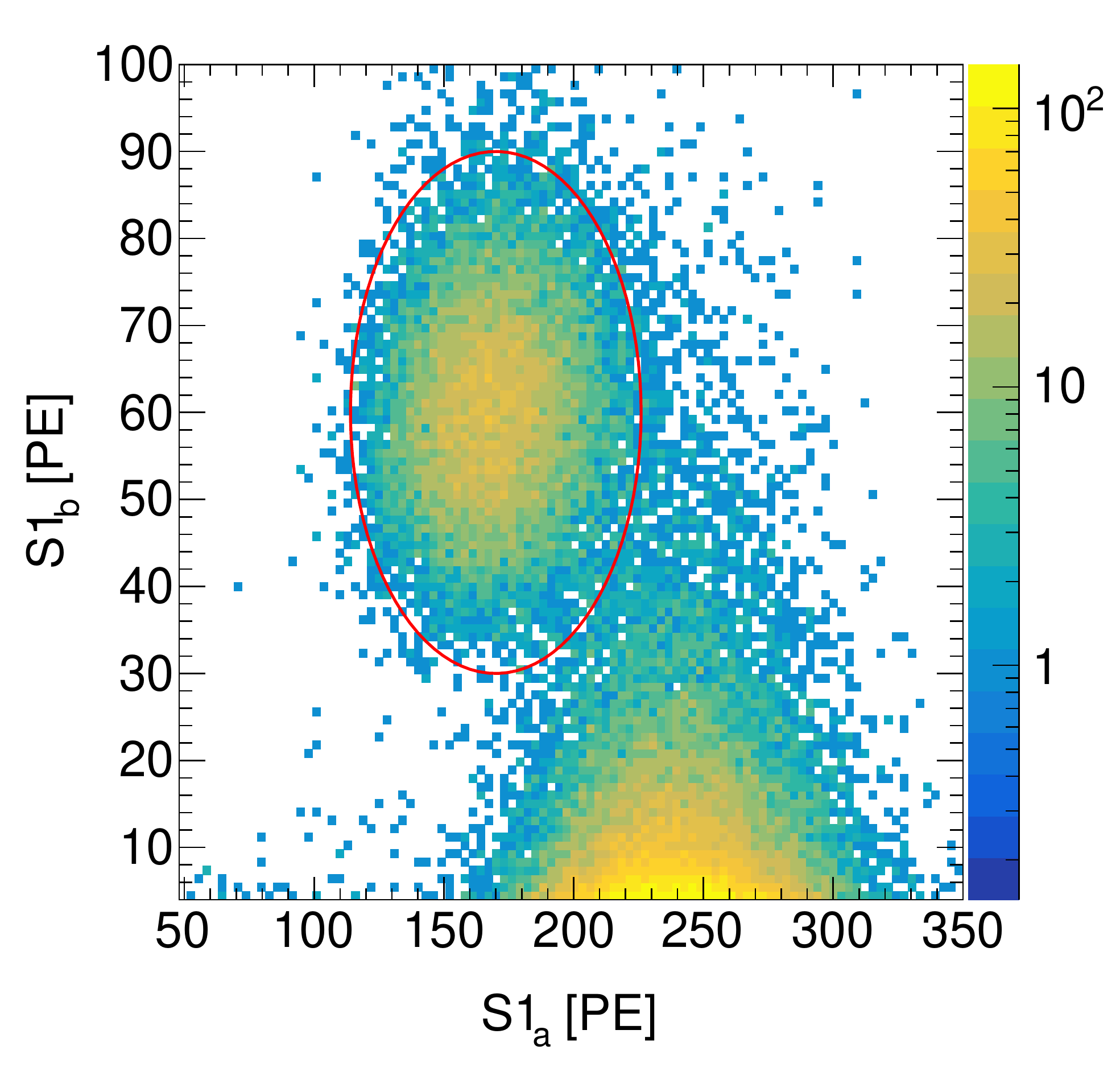}
    \label{fig:kr83mdoubleS1}
    }
    \subfigure[]{
    \includegraphics[width=0.42\textwidth]{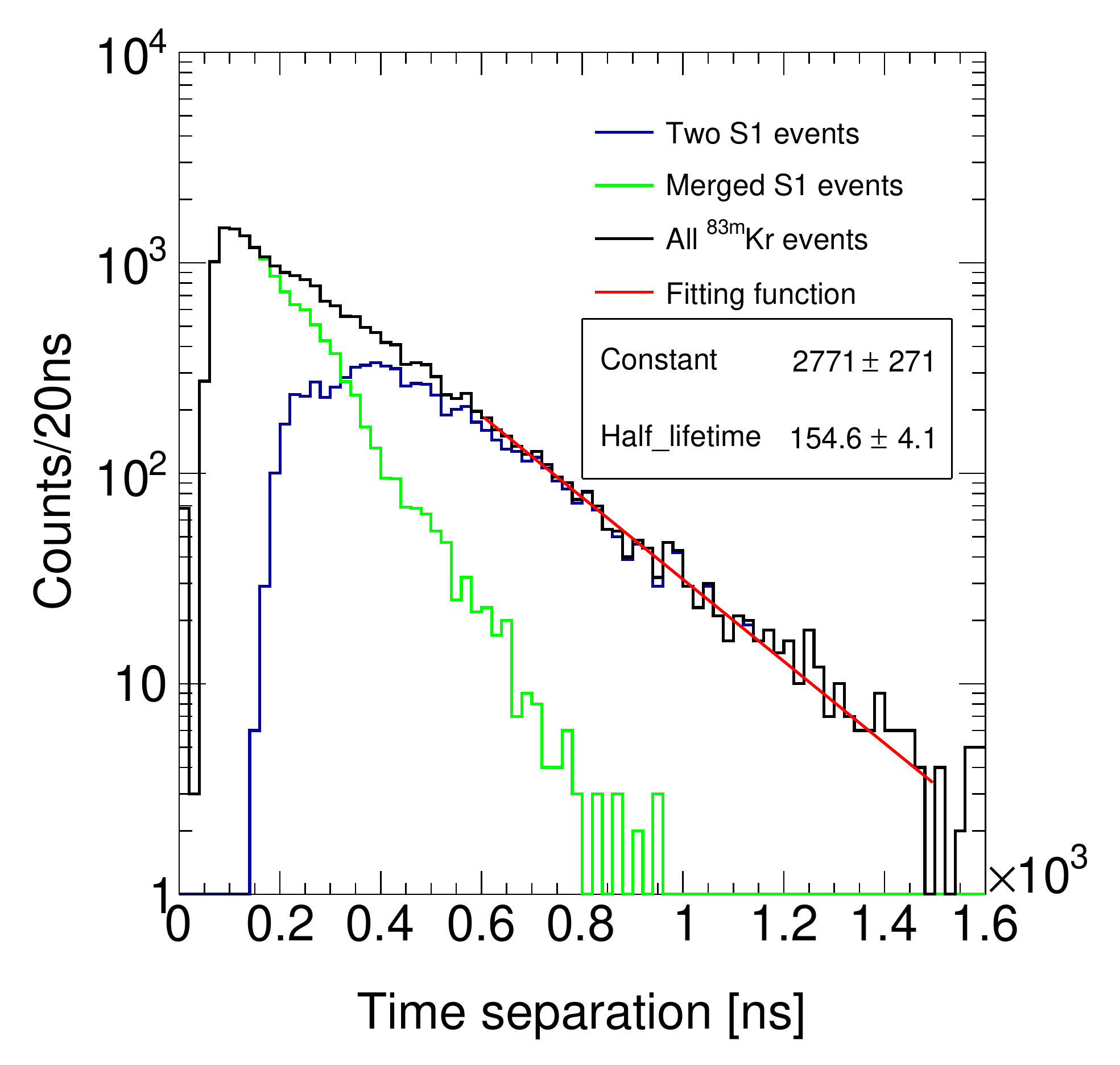}
    \label{fig:kr83mhalt}
    }
    \caption{ (a) $^{83\text{m}}$Kr $S1$ pair distribution. The horizontal axis represents the largest $S1$ in one event. The vertical axis represents the following second largest $S1$. The red circle region shows the two isolated $S1$s from the $^{83\text{m}}$Kr cascade decay.
    (b)Distribution of $^{83\text{m}}$Kr events on time separation between two peaks. 
    The horizontal axis is the peak time separation of the two signals, indicating the time feature of the second decay.
    Blue histogram represents two isolated $S1$s events within the red circle region in figure~\ref{fig:kr83mdoubleS1} 
    while the green one represents one merged $S1$ events which consists of two peaks.
    The S1 tagging algorithm affects the shapes of these two histograms. 
    However the summation of above two (black histogram) is more rational.
    Due to the inefficiency of separating two S1 peaks in small time separation region,
    only events whose delta time is within time interval from 600~ns to 1500~ns are selected to fit 
    and the fitted result is 154.6 $\pm$ 4.1~ns.
    }
\end{figure}

Figure~\ref{fig:kr83mdoubleS1} shows the event $S1$ pair distribution, 
with $S1_a$ as the largest $S1$ in the event, 
and $S1_b$ as the second largest one following $S1_a$. 
Double $S1$ events from $^{83\text{m}}$Kr are indicated as the circled region in the figure. 
The time separation of the two $S1$ peaks in figure~\ref{fig:kr83mhalt} 
obeys the exponential law 
and the fitted half life is $154.6 \pm 4.1$~ns
which is consistent with the NNDC data~\cite{DataSheet83}.

\begin{figure}[htbp]
    \centering
    \includegraphics[scale=0.4]{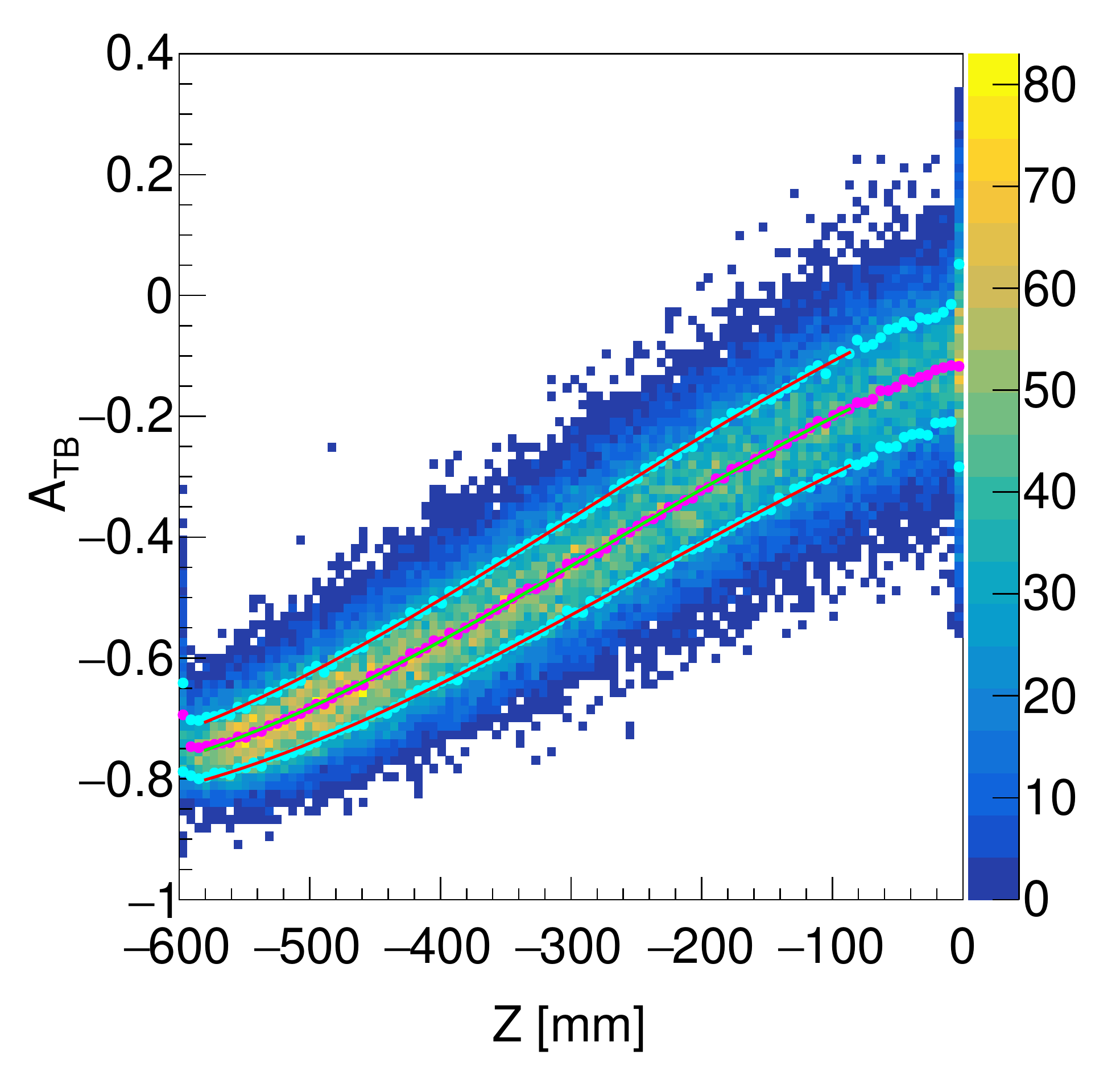}
    \caption{Distribution of $^{\text{83m}}$Kr events on normalized top-bottom ratio of $S1$ charge versus vertical position Z. Z = -600~mm and Z = 0~mm represent the cathode grid and gate mesh respectively. The 2-D histogram shows the $^{83\text{m}}$Kr event distribution. The magenta (cyan) data points are the mean ($\pm 1\sigma$) of the histogram along the vertical position axis. These relations can be fitted well as the green and red curves show.}
    \label{fig:dtTBratio}
    \vspace{\baselineskip}
\end{figure}

These $^{83\text{m}}$Kr events can be selected through the double $S1$s feature even in zero field condition, although there is no $S2$ signal. 
Without $S2$, the event position can not be reconstructed accurately.
However the normalized ratio of $S1$ charge collected in the top PMT array to that in the bottom array, 
\[
A_{\text{TB} } =  \frac{S1_{\rm top} / S1_{\rm bottom} - 1}{S1_{\rm top} / S1_{\rm bottom} + 1},
\]
reflects the event position in vertical direction Z, as demonstrated in Fig.~\ref{fig:dtTBratio} which gives A$_{\text{TB}}$-Z distribution of $^{\text{83m}}$Kr events under non-zero electric field conditions.  
A cubic polynomial function is fitted to the mean values of Z slices of the band 
to build the relationship between vertical position and normalized $S1$ top-bottom ratio $A_{\text{TB}}$,
which has no significant discrepancy at three non-zero electric field conditions.
This function is used for zero field condition to derive the detector correction factors in the vertical direction for $S1$ signals. 
Also it is obvious that the A$_{\text{TB}}$-Z relation gives an extra uncertainty on vertical position in zero field data
due to fluctuation.
And it is quantified by the $\pm~1~\sigma$ Gaussian widths of Z slices as Fig.~\ref{fig:dtTBratio} shows.

\begin{figure}[htbp]
  \centering
  \subfigure[81~V/cm]{\includegraphics[width=0.32\textwidth]{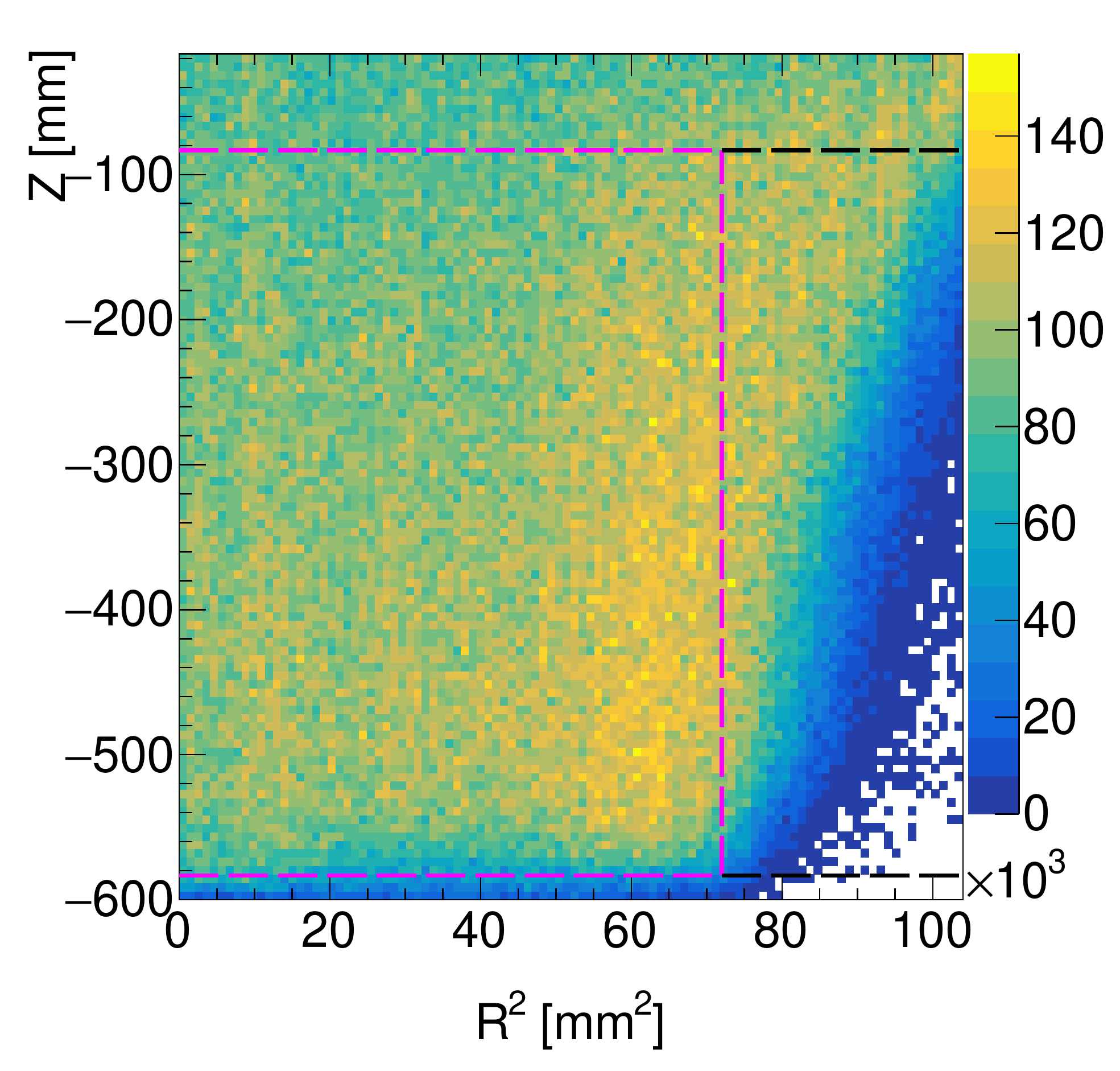}} 
  \subfigure[180~V/cm]{\includegraphics[width=0.32\textwidth]{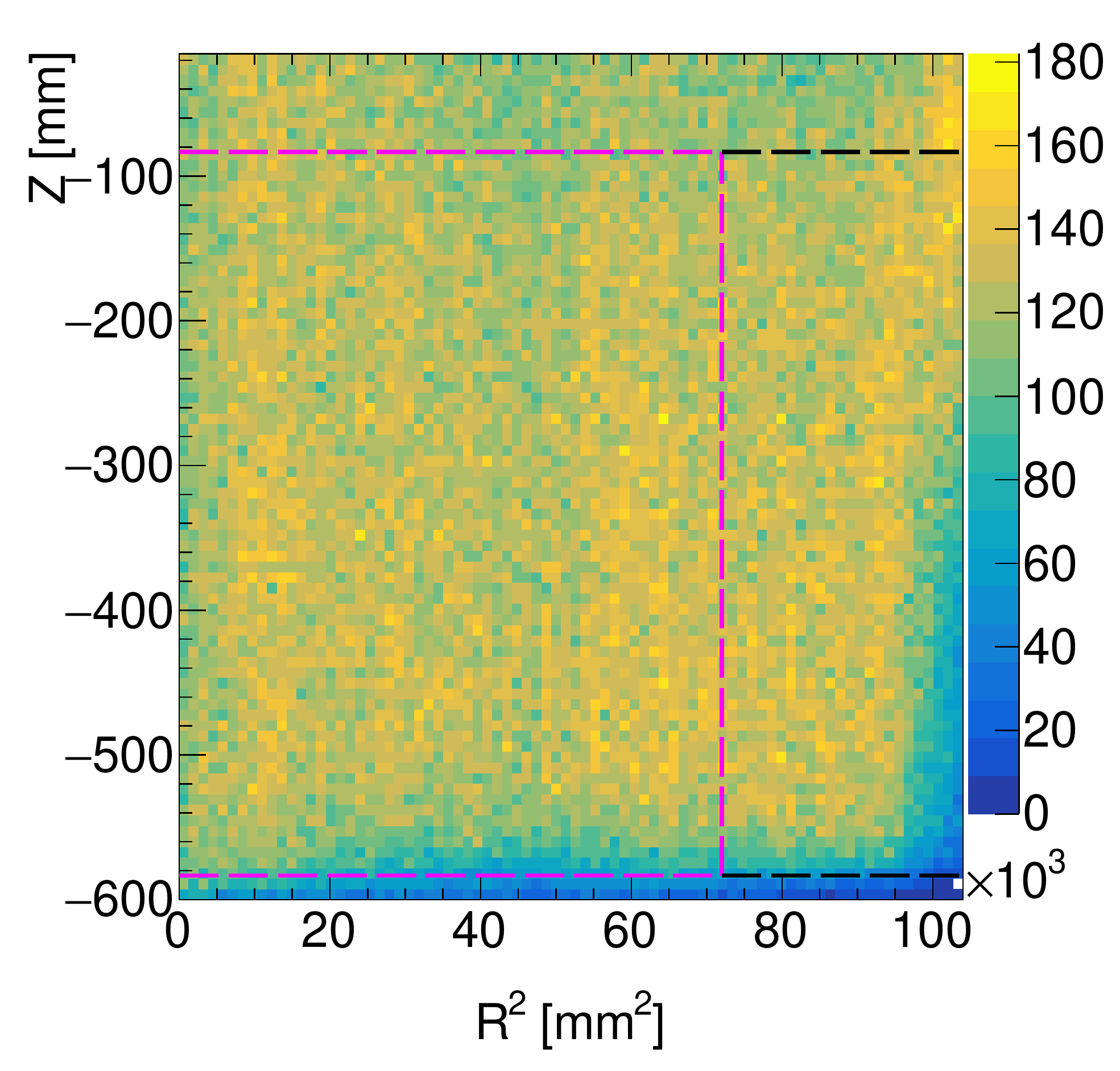}} 
  \subfigure[317~V/cm]{\includegraphics[width=0.32\textwidth]{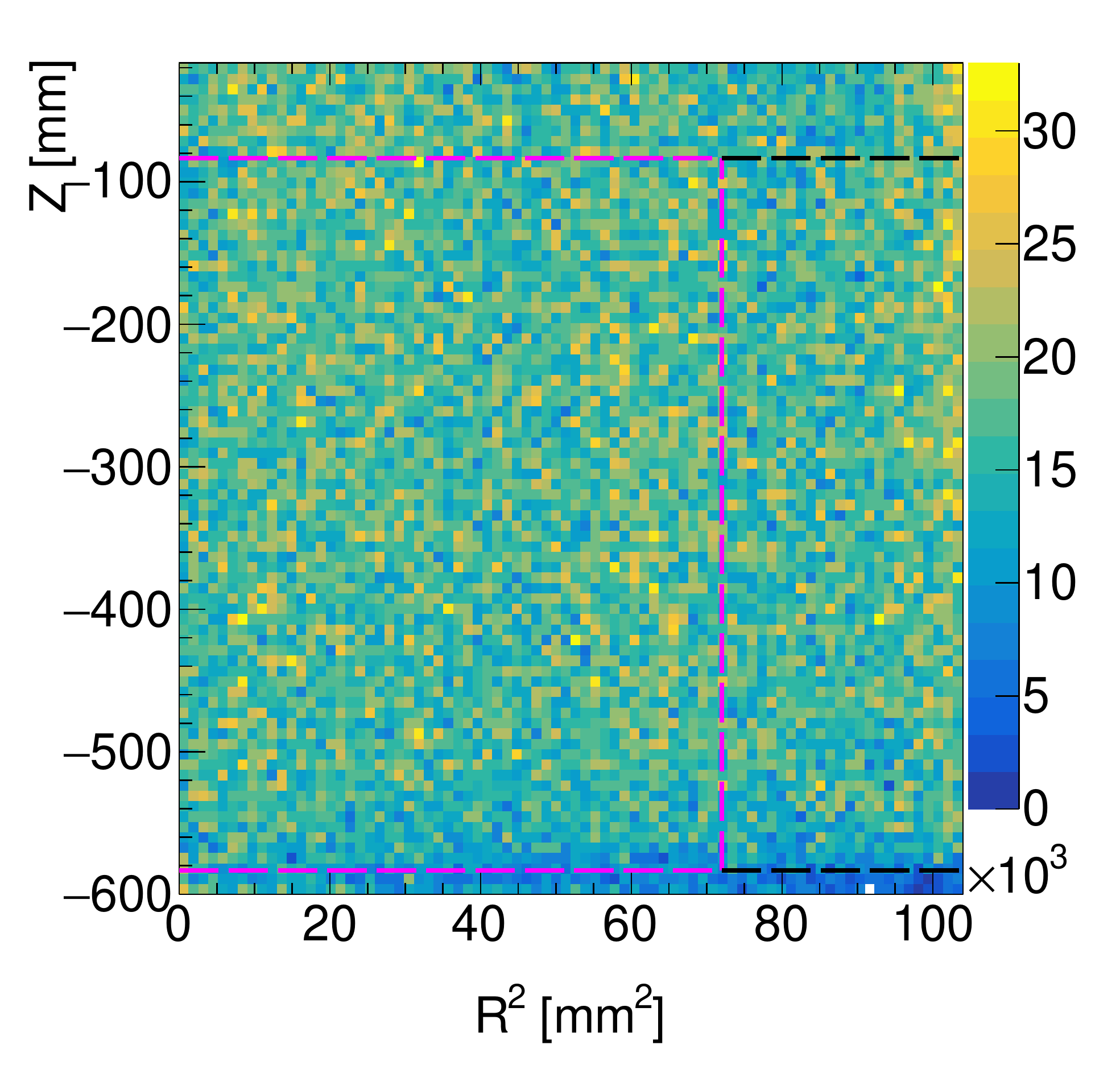}} 
  \caption{
  $^{83\text{m}}$Kr event reconstructed positions in different drift field conditions.
  The magenta and black dashed lines represent the FV and EFV boundaries, respectively.
  }
  \label{fig:krPos}
\end{figure}

Shortly after injection, the krypton atoms get distributed uniformly in the detector. 
Inside FV, our COMSOL simulation~\cite{comsol_page} predicts that the field uniformity is better than 1\%
and the electric field values uncertainties are neglected in this analysis.
However, near the edge of detector, there could exist deformation of electric field
which yields a radial component $E_r$ along with the dominant vertical component $E_z$. 
Also the inhomogeneity in this region is larger than 6\% in our simulation.
Therefore, given the large statistics of the $^{83\text{m}}$Kr events, 
the reconstructed horizontal position distribution can reflect the $E_r$ component of the drift electric field. 
The longer the drift distance is, the more distortion the reconstruction horizontal position would have.
Figure~\ref{fig:krPos} shows the reconstructed position distributions of $^{83\text{m}}$Kr events 
under various electric fields. 
The FV boundary is shown in dashed magenta lines.
And an extended fiducial volume (EFV) is confined by the dashed black lines
which removes the $R^2$ range confinement.
For events close to the detector bottom (Z~$=-600$~mm), 
the horizontal reconstructed positions shift to smaller radii. 
The signal yields of these edge events would fluctuate more
and may bias our results, 
because a stronger electric field forbids more recombination and reduces the $S1$ signal
and a weaker one has the inverse effect.
In Fig.~\ref{fig:CalculateS1}, data within FV and EFV are compared.
The $S1$ spectra of $^{83\rm{m}}$Kr data show a good consistency.
This good result comes from the low background of the selection region in figure~\ref{fig:kr83mdoubleS1} 
and indicates that edge field distortion does not affect the final results.
Thus it is convinced that the $R^2$ constrain is negligible for $^{83\rm{m}}$Kr data in zero field condition.
However, 
for activated xenon data, the mean values of the Gaussian functions are biased 
by the events from the surface (figure~\ref{fig:CalculateS1PandaX164keV} and ~\ref{fig:CalculateS1PandaX236keV}).
The consequence uncertainty will be discussed later.

\section{Results and discussion}
For a dual phase xenon detector, the light yield is an intrinsic parameter that evaluates the detector response. 
In previous sections, the $S1$ components of these monoenergetic events are selected precisely.
In this kind of large detector (646~mm in diameter and 600~mm in height), due to geometry non-uniformity, 
the signal yields can vary with event positions. 
A 3-D position-based correction is applied to improve the resolution.
The correction mapping follows Run~11 $S1$ mapping in Ref.~\cite{wang2020results}.
And all of the $S1$ charges are normalized to the mean value of $S1$ mapping inside FV.
Also the $^{131\rm{m}}$Xe mapping used for Run 9 data in Ref.~\cite{wang2020results}, 
and the different position reconstruction method introduced by Ref.~\cite{zhang2021horizontal},
are selected for uncertainty estimation in this analysis. 
Different mappings and reconstruction methods result in little measurement difference.

Besides position correction, the baseline suppression(BLS) non-linearity,  
another hardware limitation will affect the signal yield.
In a standard way, a fixed threshold is set to ignore noise which comes from baseline fluctuation. 
When the PMT's gain is high enough, 
every hit originating from physics event will be recorded 
as it is amplified larger than the threshold significantly.
However, this effect results in smaller detected $S1$ than the true $S1$ especially for the low energy signal when some of the PMTs' gains are relatively low in 2019.
A novel PMT gain calibration methodology is applied to these low gain PMTs as Ref.~\cite{wang2020results} introduced.
Following Ref.~\cite{wang2020results}, the BLS non-linearity also can be quantified and it is applied to this analysis.

\begin{figure}[htbp]
    \centering
    \subfigure[9.4 keV]{\includegraphics[width=0.3\textwidth]{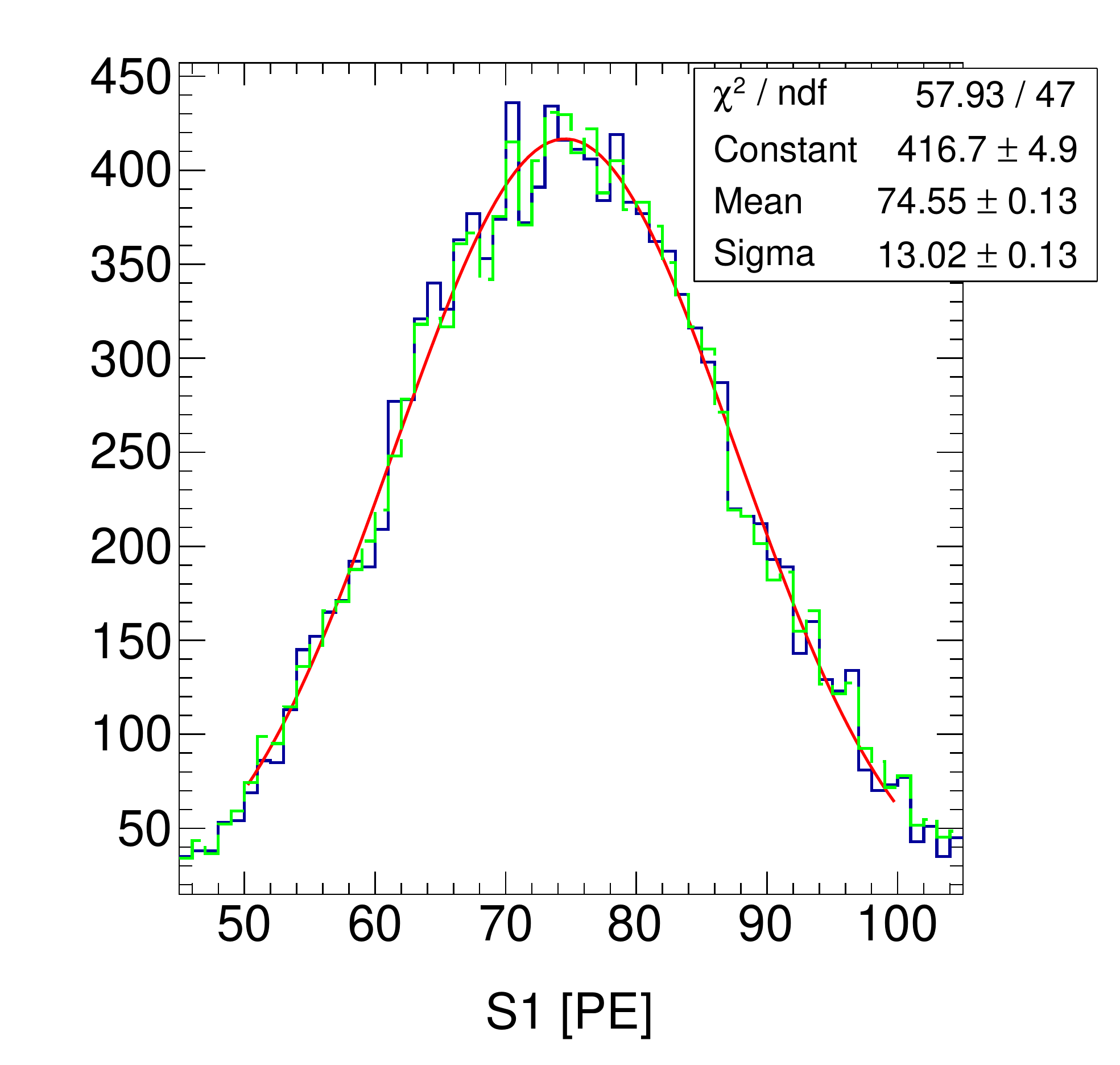}\label{fig:CalculateS1PandaX9keV}}
    \subfigure[32.1 keV]{\includegraphics[width=0.3\textwidth]{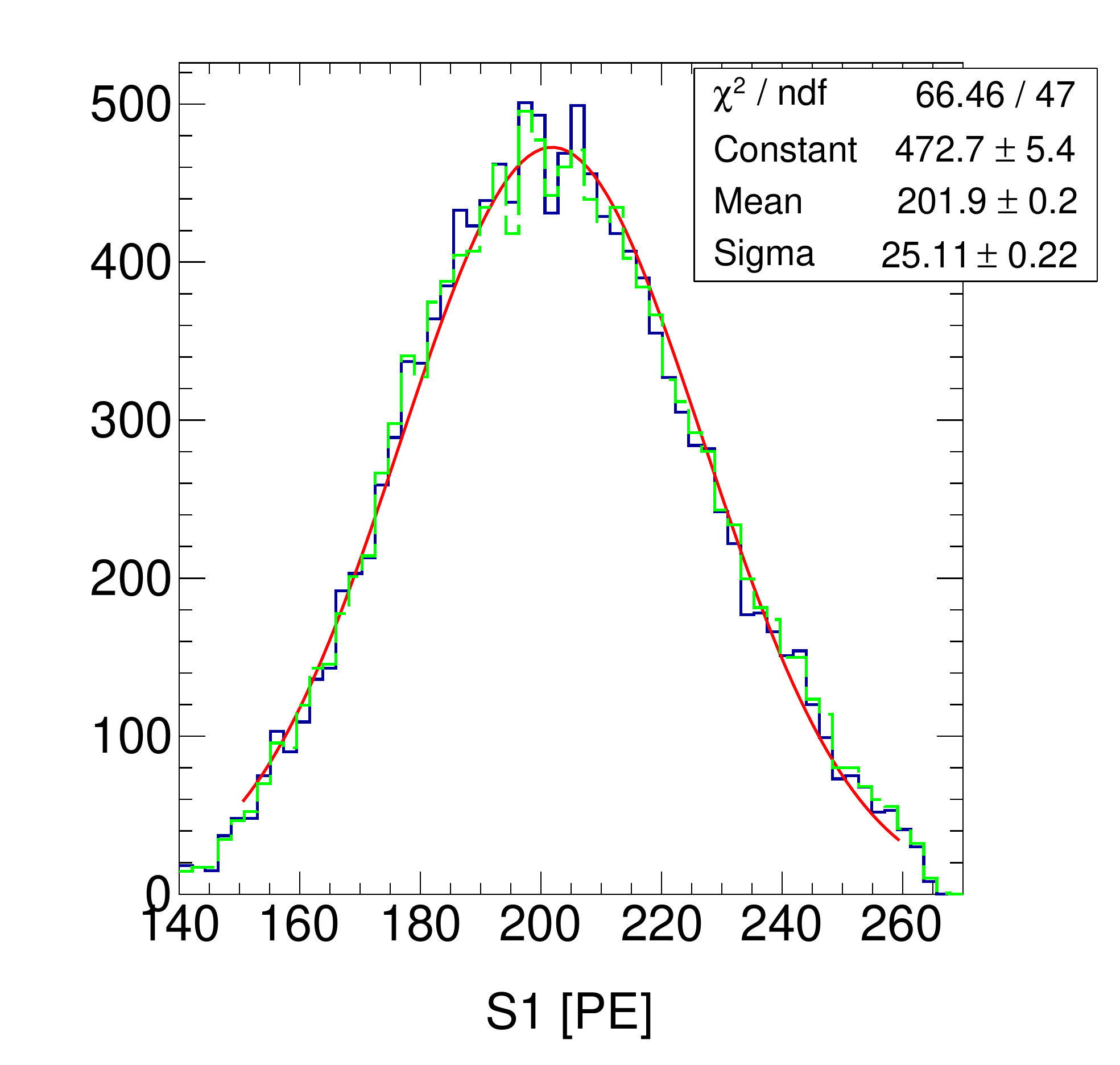}\label{fig:CalculateS1PandaX32keV}}
    \subfigure[41.5 keV]{\includegraphics[width=0.3\textwidth]{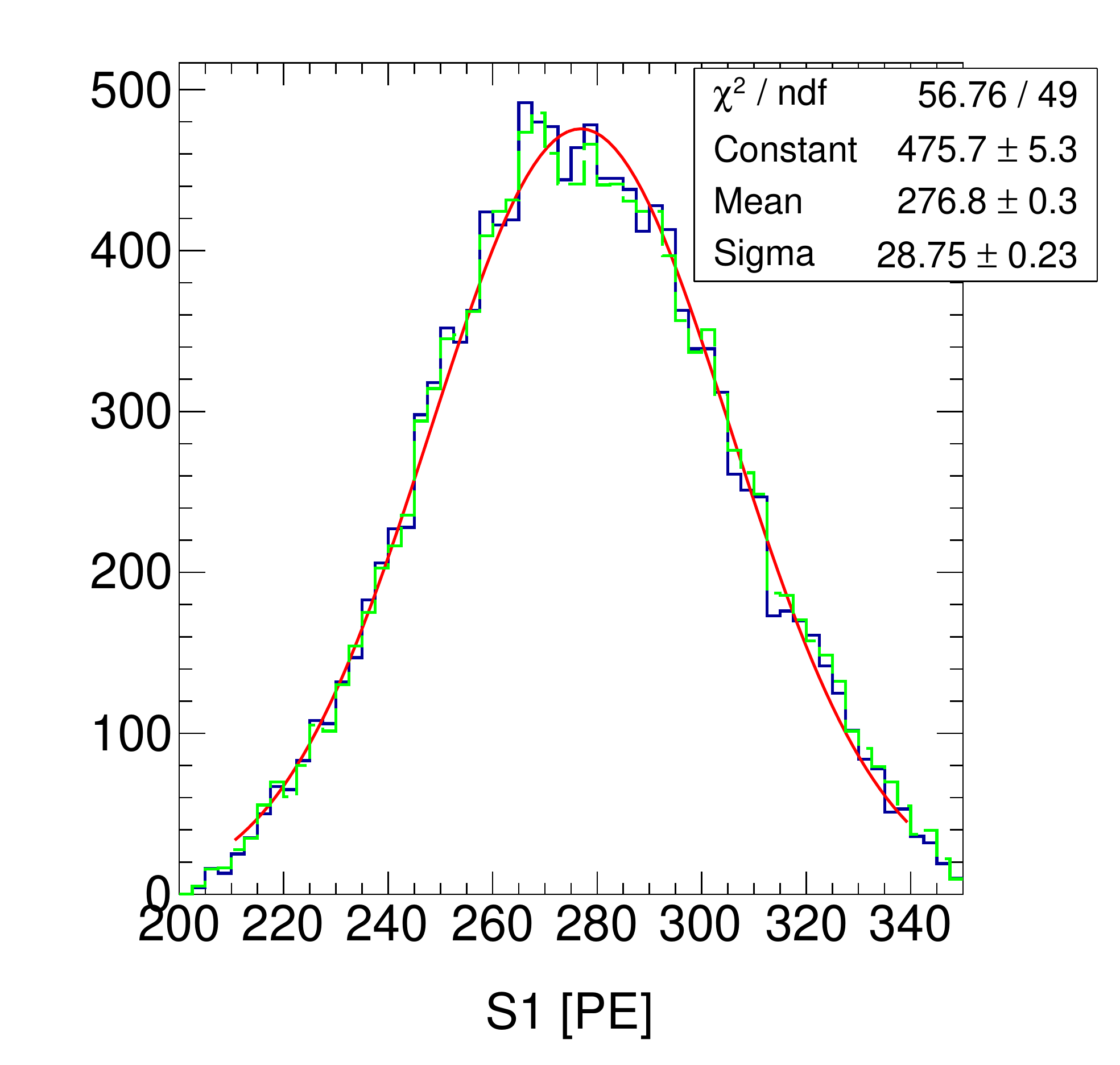}\label{fig:CalculateS1PandaX41keV}}
    \subfigure[164 keV]{\includegraphics[width=0.3\textwidth]{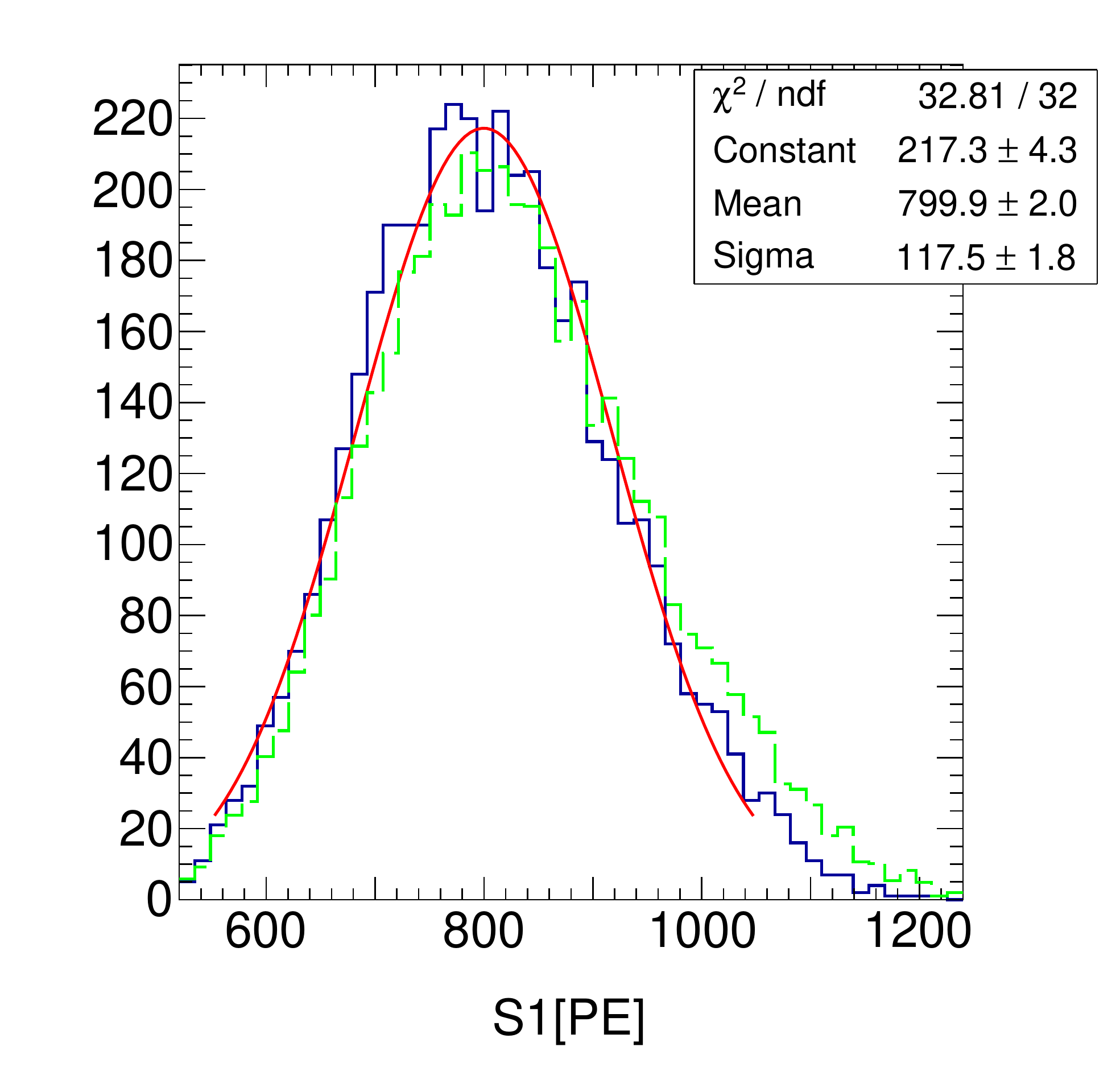}\label{fig:CalculateS1PandaX164keV}}
    \subfigure[236 keV]{\includegraphics[width=0.3\textwidth]{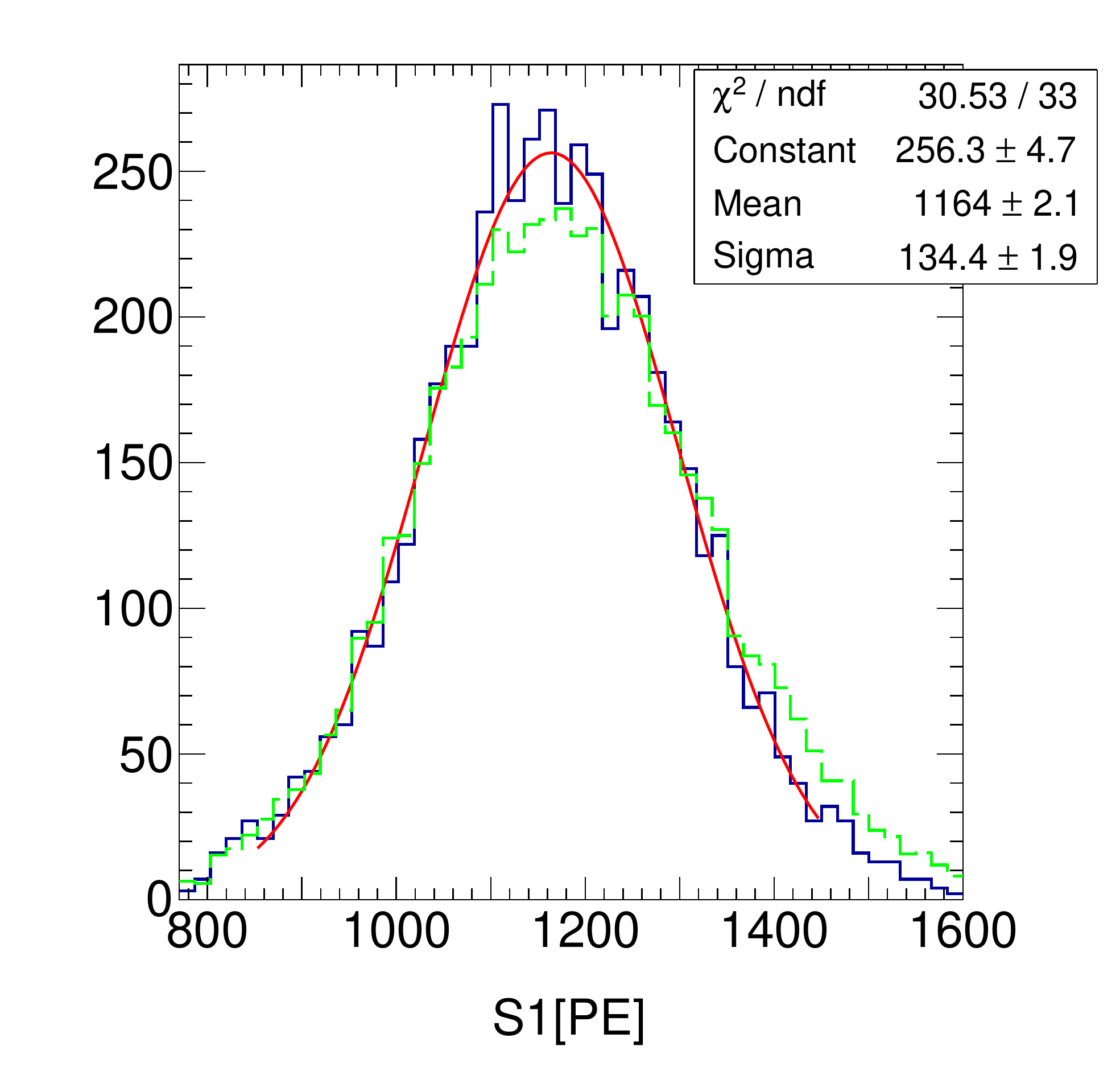}\label{fig:CalculateS1PandaX236keV}}
    \caption{
    $S1$ spectra and Gaussian function fitting at 317~V/cm electric field. 
    Blue (green) histograms are the $S1$ distributions of multiple energy points within FV (EFV). 
    The red curves are fitting functions.
    }
    \label{fig:CalculateS1}
    \vspace{\baselineskip}
\end{figure}

\begin{figure}[htbp]
    \centering
    \subfigure[9.4 keV]{\includegraphics[width=0.3\textwidth]{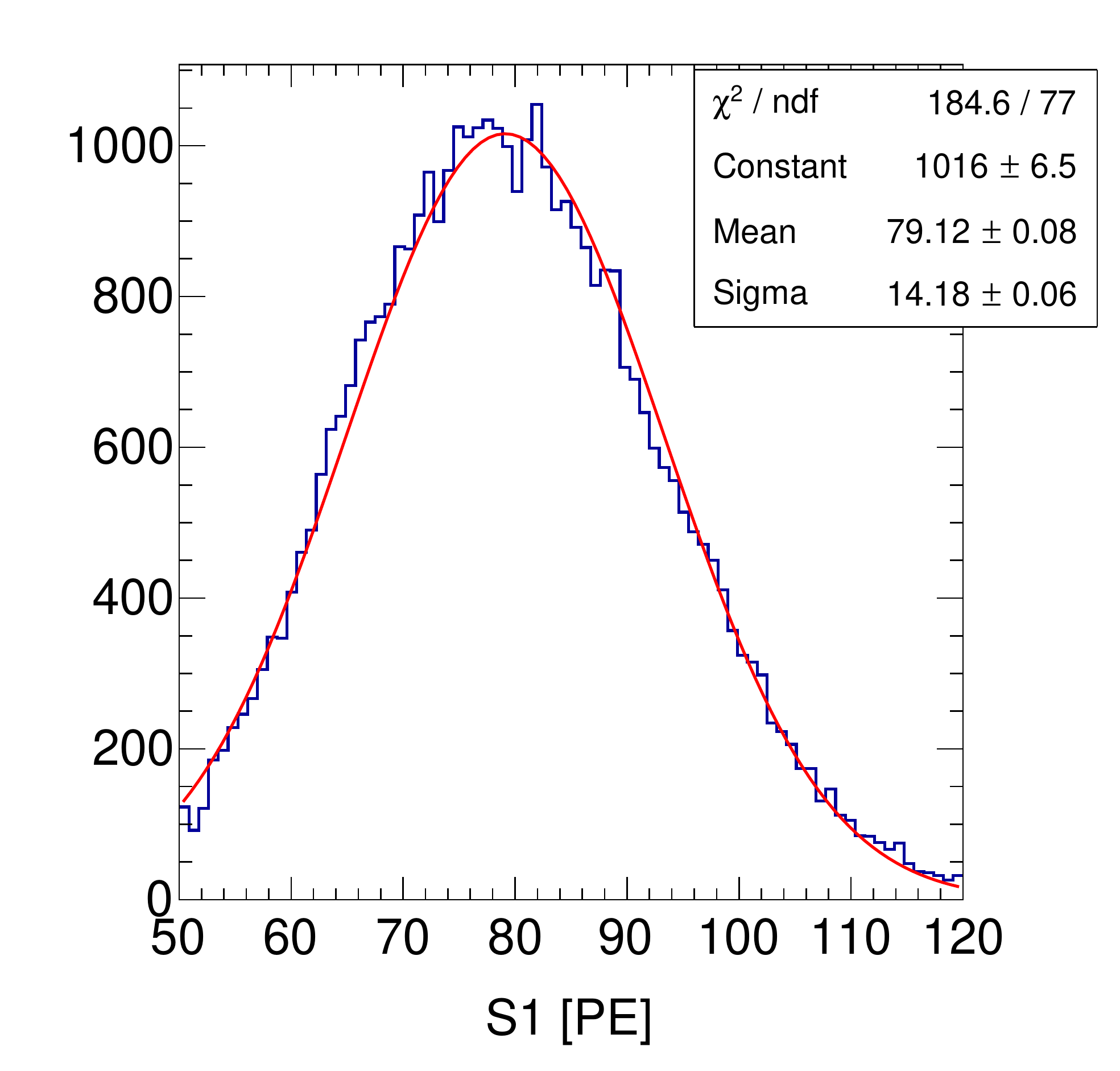}}
    \subfigure[32.1 keV]{\includegraphics[width=0.3\textwidth]{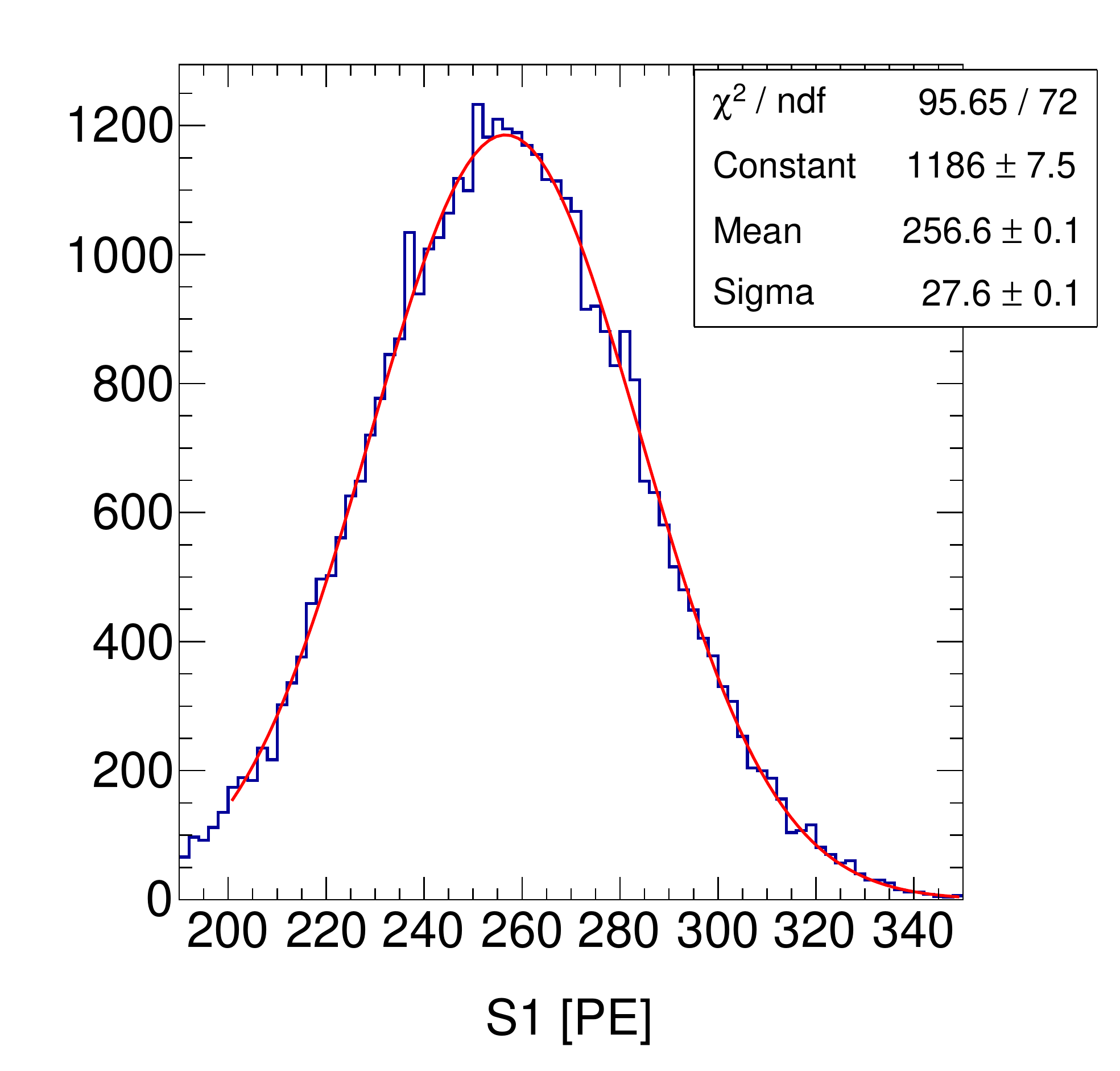}}
    \subfigure[41.5 keV]{\includegraphics[width=0.3\textwidth]{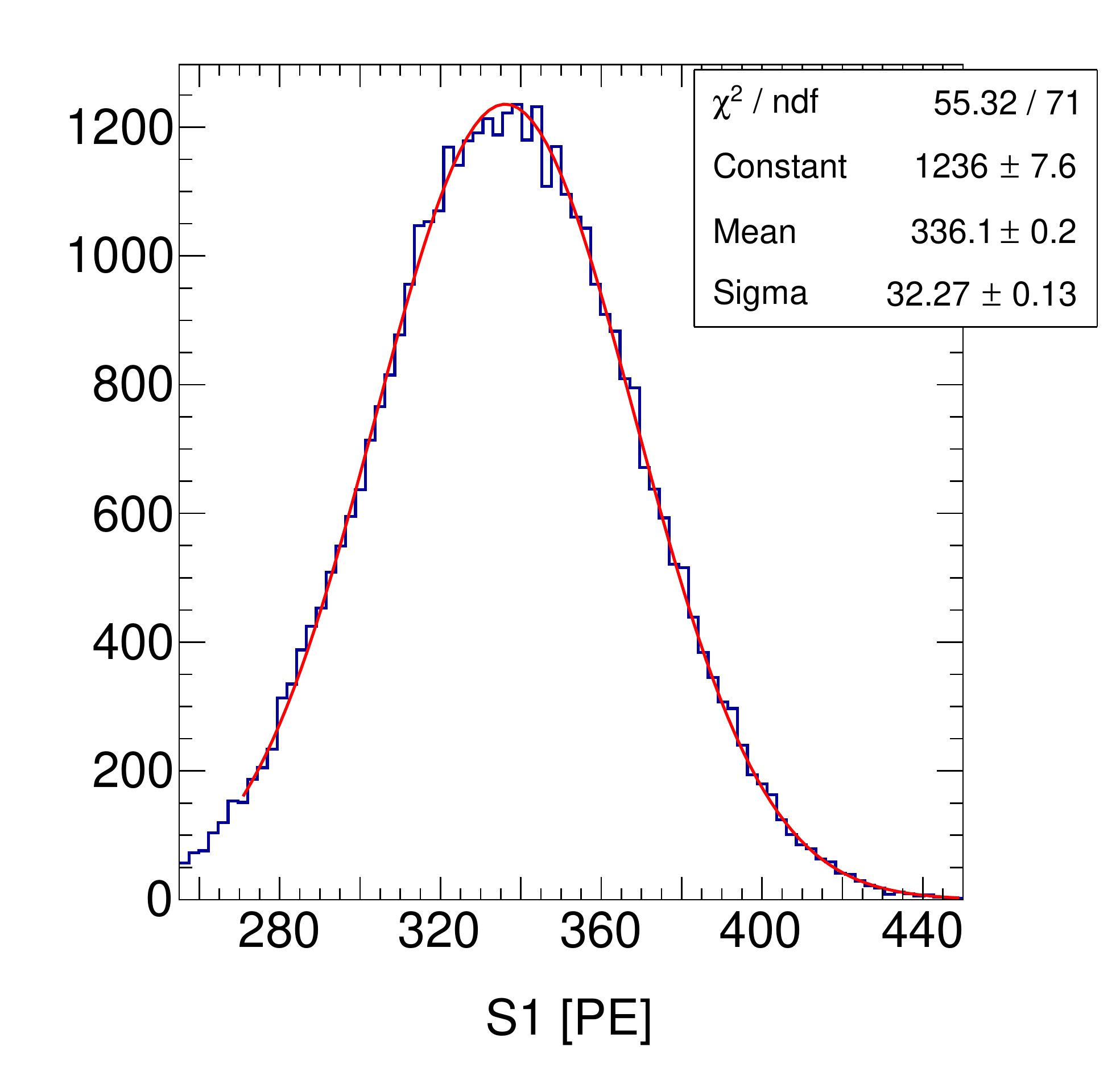}}
    \caption{\color{black} $S1$ spectra of multiple energy points and fitting functions at zero electric field.}
    \label{fig:CalculateS1_zerofield}
    \vspace{\baselineskip}
\end{figure}

After position-based correction and BLS correction, 
each $S1$ spectrum is fitted by a Gaussian function 
and the mean value of the fitting result represents the mean $S1$ response in light yield formula (\ref{eqn:LyFormula}). 
Figure~\ref{fig:CalculateS1} and \ref{fig:CalculateS1_zerofield} show 
the $S1$ spectra of various energy points at 317~V/cm and zero drift electric fields respectively. 
In association with PDE and the true energy of the decay, the light signal yield is obtained. 

The dominant systematic uncertainties originate from
the event selection methodology, Gaussian function fitting range, PDE (4\% uncertainty) and A$_{\rm{TB}}$-Z relation.
Both box and ellipse selection regions are tried for these monoenergetic ER events.
Little difference is found.
However the selection regions in Fig.~\ref{fig:ActXeS2S1} and~\ref{fig:kr83mdoubleS1} 
could be more strict and that yields 1\% uncertainty.
The $S1$ spectra of activated xenon events become non-Gaussian when surface background events contaminate it.
Then Gaussian function fitting range could yield another 1\% uncertainty.
For the zero field data,
the A$_{\rm{TB}}$-Z relation introduces extra 3\% systematic uncertainty.
Compared with all these above, the statistic uncertainty is negligible.
The results are summarized in Tab.~\ref{tab:LYresult}.

\begin{figure}[htbp]
    \centering
    \includegraphics[width=0.8\textwidth]{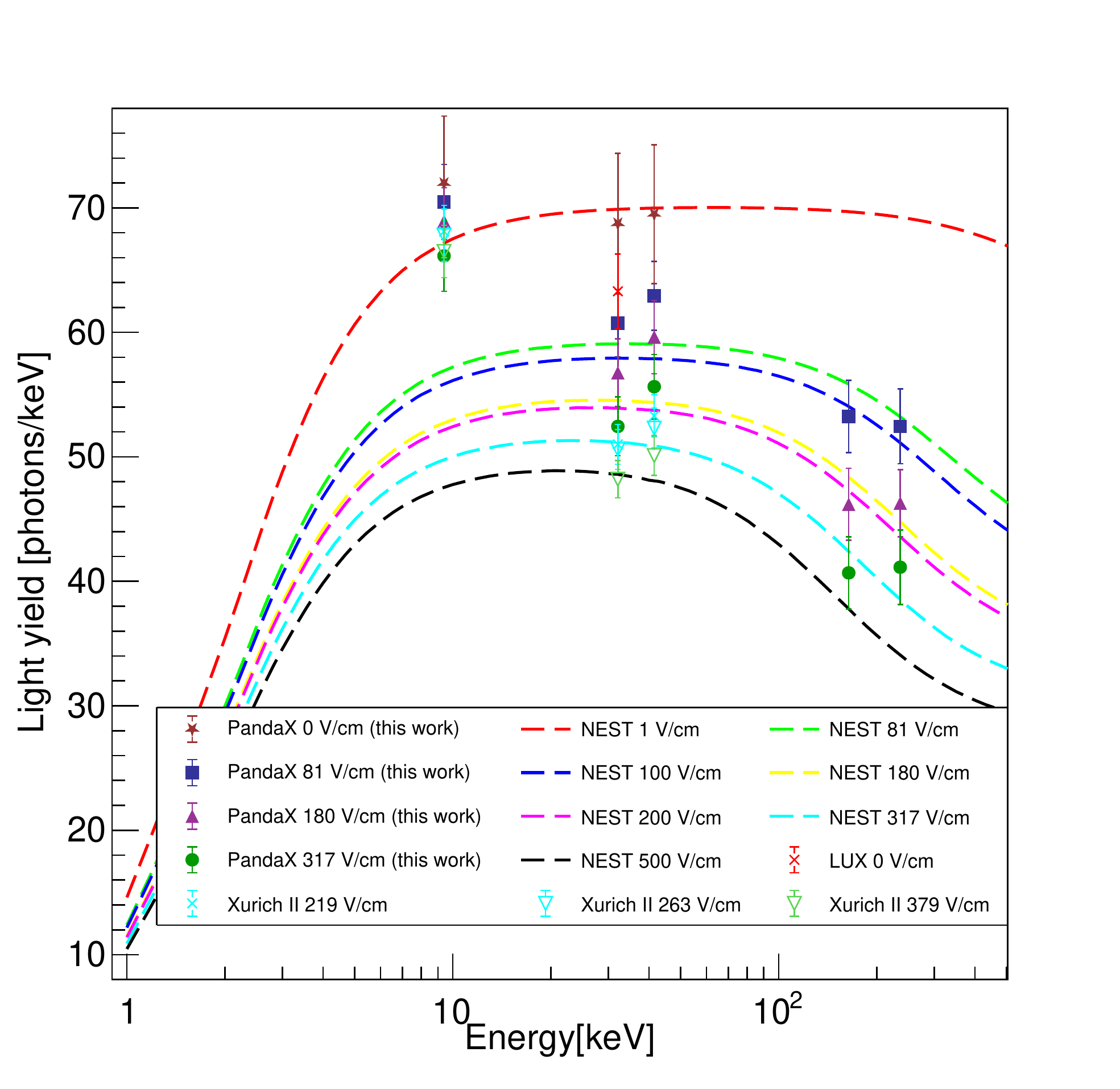}
    \caption{Light yields comparison.
    Dashed curves represent general gamma responses of NEST prediction in Ref.~\cite{Szydagis_2011}.
    The data points in this work and other experiments~\cite{osti_1257742,baudis2018dual} are also shown.
    The color set, cyan, magenta, blue, green, red and black stands for different drift electric field.
    }
    \label{fig:LYresult}
    \vspace{\baselineskip}
\end{figure}

In dual-phase noble elements based experiments, NEST model as presented in Ref.~\cite{Szydagis_2011,nest2,Szydagis_2021}, 
has been widely used for signal prediction. 
It is worthwhile measuring the light yields at several energy points to validate the NEST model. 
The measured light yields of these monoenergetic points are plotted in Fig.~\ref{fig:LYresult},
overlaid with results from other experiments~\cite{osti_1257742,baudis2018dual} and NEST model (v2.1.0) predictions~\cite{nest2}. 
The comparison shows a good consistency in general between the measurement 
and NEST model for various energies except the 9.4~keV energy point from $^{83\text{m}}$Kr cascade decay, 
which is to be discussed in the next paragraph.

\begin{table}[htbp]
\centering
\caption{\label{tab:LYresult}Light yield results, unit: photons/keV.
    These center values are calculated from the calibration events in table~\ref{tab:cali-duration}.
    Both systematic uncertainty and statistical uncertainty are considered.
    The former one includes the errors from
    selection methodology, Gaussian function fitting range, PDE,
    and A$_{\rm{TB}}$-Z relation.
    Due to the lack of $S2$ signal, these uncertainties in the 0~V/cm column are larger than others.
    }
\smallskip
\begin{tabular}{|c|cccc|}
\hline
Energy deposition (keV)  & 317~V/cm & 180~V/cm & 81~V/cm & 0~V/cm  \\
\hline
9.4  & 66.1~$\pm$~2.9 & 68.9~$\pm$~3.0 & 70.5~$\pm$~3.0 & 72.0~$\pm$~5.4 \\
32.1 & 52.4~$\pm$~2.4 & 56.8~$\pm$~2.7 & 60.7~$\pm$~2.7 & 68.7~$\pm$~5.7 \\
41.5 & 55.6~$\pm$~2.6 & 59.6~$\pm$~2.9 & 62.9~$\pm$~2.8 & 69.5~$\pm$~5.6 \\
164  & 40.7~$\pm$~2.9 & 46.2~$\pm$~2.9 & 53.2~$\pm$~2.9 & \textbf{---} \\
236  & 41.1~$\pm$~3.0 & 46.3~$\pm$~2.7 & 52.4~$\pm$~3.0 & \textbf{---} \\
\hline
\end{tabular}
\end{table}

\begin{figure}[htbp]
    \centering
    \includegraphics[width=0.7\textwidth]{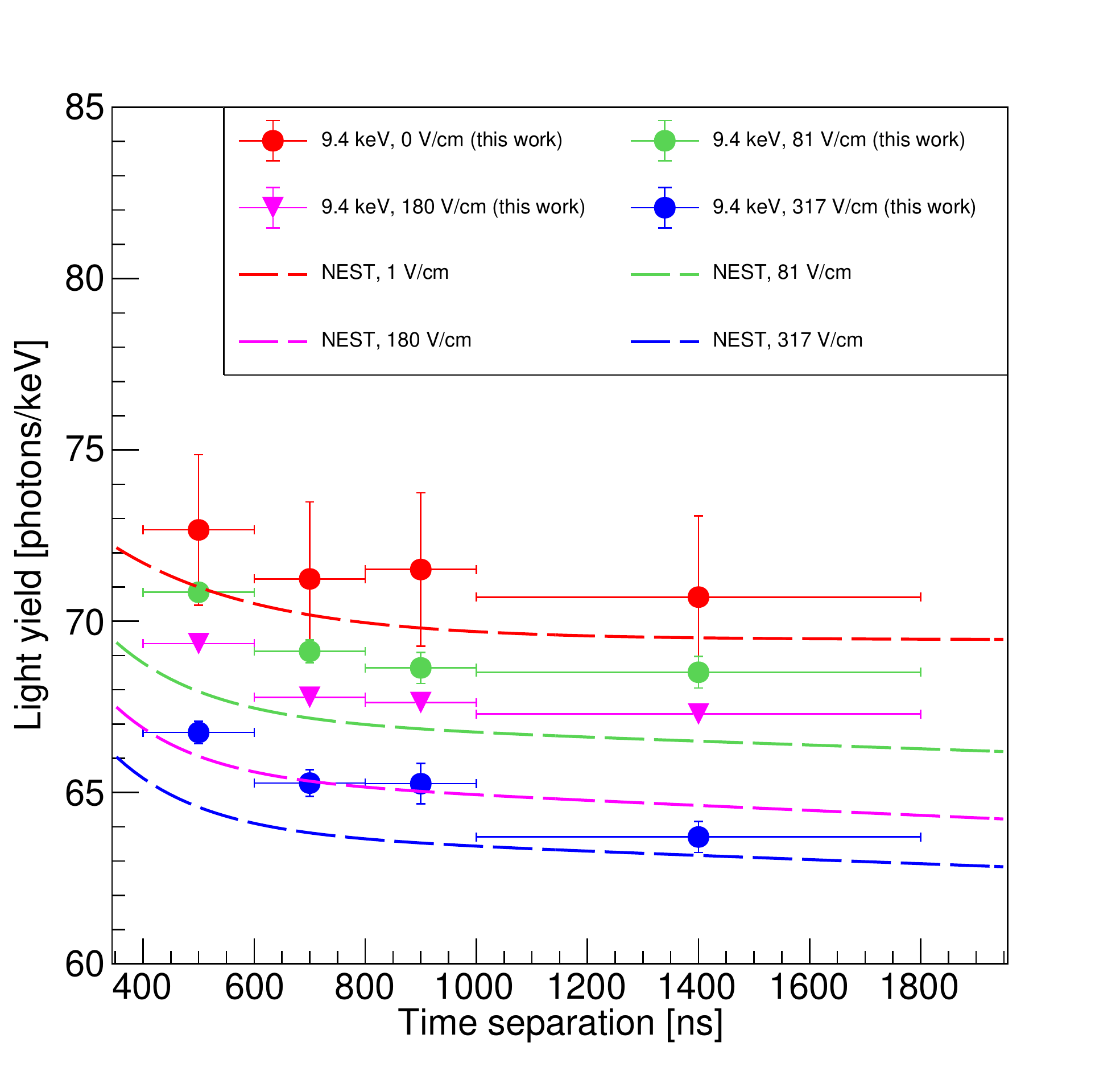}
    \caption{Light yield variation as a function of separation time for $^{83\text{m}}$Kr 9.4~keV events.
    The solid colored points are from measurement under various electric fields.
    Events with two decays time separation from 400~ns to 1800~ns are selected, 
    whose two $S1$ signals are able to be identified in the waveform. 
    The NEST model predictions for $^{83\rm{m}}$Kr are shown in dashed lines for comparison.
    The PDE uncertainty is global for the data points in this figure and it is not considered here.
    A typical decreasing dependence on the time separation is shown
    except that the red error bars are still too large to build a robust relation.
    }
    \label{fig:kr83mLYT}
    \vspace{\baselineskip}
\end{figure}

The second decay in $^{83\text{m}}$Kr cascade has quite different light yield from other energies, 
which were observed in other experiments as well~\cite{baudis:2013cca,baudis2018dual,singh2020analysis}. 
One possible interpretation is that the local high xenon ion density caused by the first decay energy deposition can enhance the ion-electron recombination rate of the surrounding second decay significantly and results in a much higher light yield than general gamma responses. 
The half-life of $^{83\text{m}}$Kr atom($J^\pi = \frac{7}{2}^+$) is 156.9~ns.
Given the large separation, 
we can expect that the second decay will have less enhancement from the first one.
Figure~\ref{fig:kr83mLYT} shows the dependence of 9.4~keV light yield on the time separation.
Also the NEST model has a similar trend as the measurement.
The dominated PDE uncertainty is not shown in this figure 
and may account for the discrepancy between data and NEST model.

\section{Summary and outlook}
After PandaX-II detector accomplishes its scientific mission~\cite{wang2020results}, two kinds of gaseous calibration sources, 
activated xenon and $^{83\text{m}}$Kr, are injected into the detector. 
In total, five energy points are reconstructed under four different electric drift field conditions. 
The signal yields are studied and show a great agreement with the NEST model.
Also, this study provides valuable experience for the next-generation experiments,
like XENONnT~\cite{XENONnT_paper}, LZ~\cite{LZTDR}, Darwin~\cite{darwin_paper},
PandaX-4T~\cite{pandax4t_sensitivity_paper,meng2021p4} and its successor~\cite{Juyal2020,Giboni2020}.

\acknowledgments
This project is supported in part by a grant from the Ministry of Science and Technology of
China (No. 2016YFA0400301), grants from National Science
Foundation of China (Nos. 12090060, 12005131, 11905128, 11925502, 11775141), 
and by Office of Science and
Technology, Shanghai Municipal Government (grant No. 18JC1410200). We thank supports from Double First Class Plan of
the Shanghai Jiao Tong University. We also thank the sponsorship from the
Chinese Academy of Sciences Center for Excellence in Particle
Physics (CCEPP), Hongwen Foundation in Hong Kong, and Tencent
Foundation in China. Finally, we thank the CJPL administration and
the Yalong River Hydropower Development Company Ltd. for
indispensable logistical support and other help.

\bibliographystyle{JHEP}
\bibliography{reference}

\providecommand{\href}[2]{#2}\begingroup\raggedright\begin{thebibliography}{10}

\bibitem{xiao2015panda}
{\scshape PandaX} collaboration, \emph{{Low-mass dark matter search results
  from full exposure of the PandaX-I experiment}},
  \href{https://doi.org/10.1103/PhysRevD.92.052004}{\emph{Phys. Rev. D}
  {\bfseries 92} (2015) 052004}
  [\href{https://arxiv.org/abs/1505.00771}{{\ttfamily 1505.00771}}].

\bibitem{Tan_2016}
{\scshape PandaX} collaboration, \emph{{Dark Matter Search Results from the
  Commissioning Run of PandaX-II}},
  \href{https://doi.org/10.1103/PhysRevD.93.122009}{\emph{Phys. Rev. D}
  {\bfseries 93} (2016) 122009}
  [\href{https://arxiv.org/abs/1602.06563}{{\ttfamily 1602.06563}}].

\bibitem{Cui:2017nnn}
{\scshape PandaX-II} collaboration, \emph{{Dark Matter Results From 54-Ton-Day
  Exposure of PandaX-II Experiment}},
  \href{https://doi.org/10.1103/PhysRevLett.119.181302}{\emph{Phys. Rev. Lett.}
  {\bfseries 119} (2017) 181302}
  [\href{https://arxiv.org/abs/1708.06917}{{\ttfamily 1708.06917}}].

\bibitem{wang2020results}
{\scshape PandaX-II} collaboration, \emph{{Results of dark matter search using
  the full PandaX-II exposure}},
  \href{https://doi.org/10.1088/1674-1137/abb658}{\emph{Chin. Phys. C}
  {\bfseries 44} (2020) 125001}
  [\href{https://arxiv.org/abs/2007.15469}{{\ttfamily 2007.15469}}].

\bibitem{luxFinal}
{\scshape LUX} collaboration, \emph{{Results from a search for dark matter in
  the complete LUX exposure}},
  \href{https://doi.org/10.1103/PhysRevLett.118.021303}{\emph{Phys. Rev. Lett.}
  {\bfseries 118} (2017) 021303}
  [\href{https://arxiv.org/abs/1608.07648}{{\ttfamily 1608.07648}}].

\bibitem{xenon1tFinal}
{\scshape XENON} collaboration, \emph{{Dark Matter Search Results from a One
  Ton-Year Exposure of XENON1T}},
  \href{https://doi.org/10.1103/PhysRevLett.121.111302}{\emph{Phys. Rev. Lett.}
  {\bfseries 121} (2018) 111302}
  [\href{https://arxiv.org/abs/1805.12562}{{\ttfamily 1805.12562}}].

\bibitem{meng2021p4}
Y.~Meng, Z.~Wang, Y.~Tao et~al., \emph{{Dark Matter Search Results from the
  PandaX-4T Commissioning Run}},
  \href{https://arxiv.org/abs/arXiv:2107.13438}{{\ttfamily arXiv:2107.13438}}.

\bibitem{xenon175wavelength}
K.~Fujii, Y.~Endo, Y.~Torigoe, S.~Nakamura, T.~Haruyama, K.~Kasami et~al.,
  \emph{High-accuracy measurement of the emission spectrum of liquid xenon in
  the vacuum ultraviolet region},
  \href{https://doi.org/https://doi.org/10.1016/j.nima.2015.05.065}{\emph{Nuclear
  Instruments and Methods in Physics Research Section A: Accelerators,
  Spectrometers, Detectors and Associated Equipment} {\bfseries 795} (2015)
  293}.

\bibitem{Plante:2011hw}
G.~Plante, E.~Aprile, R.~Budnik, B.~Choi, K.~L. Giboni, L.~W. Goetzke et~al.,
  \emph{{New Measurement of the Scintillation Efficiency of Low-Energy Nuclear
  Recoils in Liquid Xenon}},
  \href{https://doi.org/10.1103/PhysRevC.84.045805}{\emph{Phys. Rev. C}
  {\bfseries 84} (2011) 045805}
  [\href{https://arxiv.org/abs/1104.2587}{{\ttfamily 1104.2587}}].

\bibitem{ma2020internal}
W.~Ma et~al., \emph{{Internal calibration of the PandaX-II detector with radon
  gaseous sources}},
  \href{https://doi.org/10.1088/1748-0221/15/12/P12038}{\emph{JINST} {\bfseries
  15} (2020) P12038} [\href{https://arxiv.org/abs/2006.09311}{{\ttfamily
  2006.09311}}].

\bibitem{yan2021determination}
{\scshape PandaX-II} collaboration, \emph{{Determination of responses of liquid
  xenon to low energy electron and nuclear recoils using a PandaX-II
  detector}}, \href{https://doi.org/10.1088/1674-1137/abf6c2}{\emph{Chin. Phys.
  C} {\bfseries 45} (2021) 075001}
  [\href{https://arxiv.org/abs/2102.09158}{{\ttfamily 2102.09158}}].

\bibitem{aprile2018simultaneous}
E.~Aprile et~al., \emph{{Simultaneous measurement of the light and charge
  response of liquid xenon to low-energy nuclear recoils at multiple electric
  fields}}, \href{https://doi.org/10.1103/PhysRevD.98.112003}{\emph{Phys. Rev.
  D} {\bfseries 98} (2018) 112003}
  [\href{https://arxiv.org/abs/1809.02072}{{\ttfamily 1809.02072}}].

\bibitem{2019xenon1tmodel}
{\scshape XENON} collaboration, \emph{{XENON1T dark matter data analysis:
  Signal and background models and statistical inference}},
  \href{https://doi.org/10.1103/PhysRevD.99.112009}{\emph{Phys. Rev. D}
  {\bfseries 99} (2019) 112009}
  [\href{https://arxiv.org/abs/1902.11297}{{\ttfamily 1902.11297}}].

\bibitem{osti_1257742}
{\scshape LUX} collaboration, \emph{{Tritium calibration of the LUX dark matter
  experiment}}, \href{https://doi.org/10.1103/PhysRevD.93.072009}{\emph{Phys.
  Rev. D} {\bfseries 93} (2016) 072009}
  [\href{https://arxiv.org/abs/1512.03133}{{\ttfamily 1512.03133}}].

\bibitem{akerib2017signal}
{\scshape LUX} collaboration, \emph{{Signal yields, energy resolution, and
  recombination fluctuations in liquid xenon}},
  \href{https://doi.org/10.1103/PhysRevD.95.012008}{\emph{Phys. Rev. D}
  {\bfseries 95} (2017) 012008}
  [\href{https://arxiv.org/abs/1610.02076}{{\ttfamily 1610.02076}}].

\bibitem{lux2016dd}
{\scshape LUX} collaboration, \emph{{Low-energy (0.7-74 keV) nuclear recoil
  calibration of the LUX dark matter experiment using D-D neutron scattering
  kinematics}},  \href{https://arxiv.org/abs/1608.05381}{{\ttfamily
  1608.05381}}.

\bibitem{Szydagis_2021}
M.~Szydagis et~al., \emph{{A Review of Basic Energy Reconstruction Techniques
  in Liquid Xenon and Argon Detectors for Dark Matter and Neutrino Physics
  Using NEST}},
  \href{https://doi.org/10.3390/instruments5010013}{\emph{Instruments}
  {\bfseries 5} (2021) 13} [\href{https://arxiv.org/abs/2102.10209}{{\ttfamily
  2102.10209}}].

\bibitem{akerib2015radiogenic}
D.~S. Akerib et~al., \emph{{Radiogenic and Muon-Induced Backgrounds in the LUX
  Dark Matter Detector}},
  \href{https://doi.org/10.1016/j.astropartphys.2014.07.009}{\emph{Astropart.
  Phys.} {\bfseries 62} (2015) 33}
  [\href{https://arxiv.org/abs/1403.1299}{{\ttfamily 1403.1299}}].

\bibitem{kr83myale}
L.~W. Kastens, S.~B. Cahn, A.~Manzur and D.~N. McKinsey, \emph{{Calibration of
  a Liquid Xenon Detector with Kr-83m}},
  \href{https://doi.org/10.1103/PhysRevC.80.045809}{\emph{Phys. Rev. C}
  {\bfseries 80} (2009) 045809}
  [\href{https://arxiv.org/abs/0905.1766}{{\ttfamily 0905.1766}}].

\bibitem{kr83mzurich}
A.~Manalaysay et~al., \emph{{Spatially uniform calibration of a liquid xenon
  detector at low energies using 83m-Kr}},
  \href{https://doi.org/10.1063/1.3436636}{\emph{Rev. Sci. Instrum.} {\bfseries
  81} (2010) 073303} [\href{https://arxiv.org/abs/0908.0616}{{\ttfamily
  0908.0616}}].

\bibitem{LUXkr83m}
{\scshape LUX} collaboration, \emph{{$^{83\textrm{m}}$Kr calibration of the
  2013 LUX dark matter search}},
  \href{https://doi.org/10.1103/PhysRevD.96.112009}{\emph{Phys. Rev. D}
  {\bfseries 96} (2017) 112009}
  [\href{https://arxiv.org/abs/1708.02566}{{\ttfamily 1708.02566}}].

\bibitem{zhangdan83mKr}
D.~Zhang et~al., \emph{{Rb83/Kr83m production and cross-section measurement
  with 3.4 MeV and 20 MeV proton beams}},
  \href{https://arxiv.org/abs/arXiv:2102.02490}{{\ttfamily arXiv:2102.02490}}.

\bibitem{DataSheet83}
E.~McCutchan, \emph{{Nuclear Data Sheets for A = 83}},
  \href{https://doi.org/https://doi.org/10.1016/j.nds.2015.02.002}{\emph{Nuclear
  Data Sheets} {\bfseries 125} (2015) 201}.

\bibitem{comsol_page}
COMSOL, ``{COMSOL Multiphysics}.'' \url{https://cn.comsol.com}.

\bibitem{zhang2021horizontal}
D.~Zhang, A.~Tan, A.~Abdukerim, W.~Chen, X.~Chen, Y.~Chen et~al.,
  \emph{{Horizontal Position Reconstruction in PandaX-II}},
  \href{https://arxiv.org/abs/arXiv:2106.08380}{{\ttfamily arXiv:2106.08380}}.

\bibitem{Szydagis_2011}
M.~Szydagis et~al., \emph{{NEST: A Comprehensive Model for Scintillation Yield
  in Liquid Xenon}},
  \href{https://doi.org/10.1088/1748-0221/6/10/P10002}{\emph{JINST} {\bfseries
  6} (2011) P10002} [\href{https://arxiv.org/abs/1106.1613}{{\ttfamily
  1106.1613}}].

\bibitem{nest2}
M.~Szydagis et~al., \emph{Noble element simulation technique},  June, 2020.
\newblock 10.5281/zenodo.3905382.

\bibitem{baudis2018dual}
L.~Baudis, Y.~Biondi, C.~Capelli, M.~Galloway, S.~Kazama, A.~Kish et~al.,
  \emph{{A Dual-phase Xenon TPC for Scintillation and Ionisation Yield
  Measurements in Liquid Xenon}},
  \href{https://doi.org/10.1140/epjc/s10052-018-5801-5}{\emph{Eur. Phys. J. C}
  {\bfseries 78} (2018) 351}
  [\href{https://arxiv.org/abs/1712.08607}{{\ttfamily 1712.08607}}].

\bibitem{baudis:2013cca}
L.~Baudis, H.~Dujmovic, C.~Geis, A.~James, A.~Kish, A.~Manalaysay et~al.,
  \emph{{Response of liquid xenon to Compton electrons down to 1.5 keV}},
  \href{https://doi.org/10.1103/PhysRevD.87.115015}{\emph{Phys. Rev. D}
  {\bfseries 87} (2013) 115015}
  [\href{https://arxiv.org/abs/1303.6891}{{\ttfamily 1303.6891}}].

\bibitem{singh2020analysis}
A.~G. Singh et~al., \emph{{Analysis of $^{83m}$Kr prompt scintillation signals
  in the PIXeY detector}},
  \href{https://doi.org/10.1088/1748-0221/15/01/P01023}{\emph{JINST} {\bfseries
  15} (2020) P01023} [\href{https://arxiv.org/abs/1911.03999}{{\ttfamily
  1911.03999}}].

\bibitem{XENONnT_paper}
{\scshape XENON} collaboration, \emph{{Projected WIMP sensitivity of the
  XENONnT dark matter experiment}},
  \href{https://doi.org/10.1088/1475-7516/2020/11/031}{\emph{JCAP} {\bfseries
  11} (2020) 031} [\href{https://arxiv.org/abs/2007.08796}{{\ttfamily
  2007.08796}}].

\bibitem{LZTDR}
B.~J. Mount et~al., \emph{{LUX-ZEPLIN (LZ) Technical Design Report}},
  \href{https://arxiv.org/abs/1703.09144}{{\ttfamily 1703.09144}}.

\bibitem{darwin_paper}
{\scshape DARWIN} collaboration, \emph{{DARWIN \textendash{} a next-generation
  liquid xenon observatory for dark matter and neutrino physics}},
  \href{https://doi.org/10.22323/1.395.0548}{\emph{PoS} {\bfseries ICRC2021}
  (2021) 548}.

\bibitem{pandax4t_sensitivity_paper}
{\scshape PandaX} collaboration, \emph{{Dark matter direct search sensitivity
  of the PandaX-4T experiment}},
  \href{https://doi.org/10.1007/s11433-018-9259-0}{\emph{Sci. China Phys. Mech.
  Astron.} {\bfseries 62} (2019) 31011}
  [\href{https://arxiv.org/abs/1806.02229}{{\ttfamily 1806.02229}}].

\bibitem{Juyal2020}
P.~Juyal, K.-L. Giboni, X.-D. Ji and J.-L. Liu, \emph{{On proportional
  scintillation in very large liquid xenon detectors}},
  \href{https://doi.org/10.1007/s41365-020-00797-4}{\emph{Nucl. Sci. Tech.}
  {\bfseries 31} (2020) 93} [\href{https://arxiv.org/abs/1910.13160}{{\ttfamily
  1910.13160}}].

\bibitem{Giboni2020}
K.~L. Giboni, P.~Juyal, E.~Aprile, Y.~Zhang and J.~Naganoma, \emph{{A
  LN$_{2}$-based cooling system for a next-generation liquid xenon dark matter
  detector}}, \href{https://doi.org/10.1007/s41365-020-00786-7}{\emph{Nucl.
  Sci. Tech.} {\bfseries 31} (2020) 76}
  [\href{https://arxiv.org/abs/1909.09698}{{\ttfamily 1909.09698}}].

\end{thebibliography}\endgroup

\end{document}